\documentclass{aa}  
\usepackage{graphicx}
\usepackage[version=4]{mhchem}
\usepackage{units}
\usepackage{amsmath,amssymb}
\usepackage[utf8]{inputenc}
\usepackage[T1]{fontenc}
\usepackage{xcolor}
\usepackage{appendix}
\usepackage{booktabs}
\usepackage{multirow}
\usepackage{graphicx}
\usepackage{txfonts}
\begin{document}

   \title{Resonant infrared irradiation of \ce{CO} and \ce{CH3OH} interstellar ices}


   \author{J. C. Santos
          \inst{1}
          \and
          K.-J. Chuang\inst{1}
          \and
          J. G. M. Schrauwen\inst{2}
          \and
          A. Traspas Muiña\inst{3}
          \and
          J. Zhang\inst{3}
          \and
          H. M. Cuppen\inst{4}
          \and
          B. Redlich\inst{2}
          \and
          H. Linnartz\inst{1}
          \and
          S. Ioppolo\inst{5, 3}
          }


            \institute{Laboratory for Astrophysics, Leiden Observatory, Leiden University, PO Box 9513, 2300 RA Leiden, The Netherlands\\
                \email{santos@strw.leidenuniv.nl}
         \and FELIX Laboratory, Radboud University, Toernooiveld 7, 6525 ED Nijmegen, The Netherlands
         \and School of Electronic Engineering and Computer Science, Queen Mary University of London, London E1 4NS, UK
         \and Institute for Molecules and Materials, Radboud University, Heyendaalseweg 135, 6525 AJ Nijmegen, The Netherlands
         \and Center for Interstellar Catalysis, Department of Physics and Astronomy, Aarhus University, Ny Munkegade 120, Aarhus C 8000, Denmark
                }


 
  \abstract
   {Solid-phase photo-processes involving icy dust grains greatly affect the chemical evolution of the interstellar medium by leading to the formation of complex organic molecules and by inducing photodesorption. So far, the focus of laboratory studies has been mainly on the impact of energetic ultraviolet (UV) photons on ices, but direct vibrational excitation by infrared (IR) photons is expected to influence the morphology and content of interstellar ices as well. However, little is still known about the mechanisms through which this excess vibrational energy is dissipated, and its implications on the structure and ice photochemistry.} 
   {In this work, we present a systematic investigation of the behavior of interstellar relevant \ce{CO} and \ce{CH3OH} ice analogues upon resonant excitation of vibrational modes using tunable infrared radiation, leading to both the quantification of infrared-induced photodesorption and insights in the impact of vibrational energy dissipation on ice morphology.}
   {We utilize an ultrahigh vacuum setup at cryogenic temperatures to grow pure \ce{CO} and \ce{CH3OH} ices, as well as mixtures of the two. We expose the ices to intense, near-monochromatic mid-infrared free-electron-laser radiation using the LISA end-station at the FELIX free electron laser facility to selectively excite the species. Changes to the ice are monitored by means of reflection-absorption infrared spectroscopy combined with quadrupole mass-spectrometry. The methods also allow to characterize the photodesorption efficiency.}
   {The dissipation of vibrational energy is observed to be highly dependent on the excited mode and the chemical environment of the ice. All amorphous ices undergo some degree of restructuring towards a more organized configuration upon on-resonance irradiation. Moreover, IR-induced photodesorption is observed to occur for both pure \ce{CO} and \ce{CH3OH} ices, with interstellar photodesorption efficiencies of the order of 10 molecules cm$^{-2}$ s$^{-1}$---i.e., comparable to or higher than UV-induced counterparts. Indirect photodesorption of \ce{CO} upon vibrational excitation of \ce{CH3OH} in ice mixtures is also observed to occur, particularly in environments rich in methanol. The astrochemical implications of these IR-induced phenomena are discussed.}
   {}


   \maketitle
%
\section{Introduction}\label{sec:intro}

Interstellar icy dust grains are continuously exposed to both energetic and non-energetic triggers that initiate solid-state processes, such as chemical reactions. The consequent association of induced fragments---e.g., in exothermic reactions---has been reported to result in larger molecules containing excess energy. The ejection of solid-state chemical products to the gas phase through so-called "reactive desorption" and "photo-induced chemical desorption" is a direct consequence of it \citep{Minissale2014, Minissale2016, Munoz-Caro2018, Chuang2018, Oba2018}. This extra energy, which can be in vibrational form, must in turn be dissipated within the ice itself. Understanding the mechanisms through which this energy is dissipated, and its consequences to the content and morphology of the ice, is of special relevance to construct accurate astrochemical models. In the last decades the focus has been on ultraviolet photons, as these are able to break chemical bonds. However, also impacting infrared photons are expected to have substantial effects, specifically in  dense molecular clouds. The infrared interstellar radiation field (ISRF) is most intense in the inner galaxy, and is generated both by stellar emission (mainly for $\lambda\lesssim8$ $\mu$m) and by re-emission from dust grains (dominant for $\lambda\gtrsim8$ $\mu$m). The former is accounted for by a combination of blackbody radiation fields, while the latter requires a model of the interstellar dust such as described by \cite{Draine2001} and \cite{Li2001}. Within clouds, the integrated flux of secondary UV photons resulting from the excitation of \ce{H2} upon interaction with cosmic rays is expected to be of the order of $\sim10^4$ photons cm$^{-2}$ s$^{-1}$ \citep{CecchiPestellini1992}, but that of IR photons with wavelengths in the range of $1-10$ $\mu$m is currently estimated to be substantially higher and of the order of $\gtrsim10^9$ photons cm$^{-2}$ s$^{-1}$ (e.g., \citealt{Mathis1983, Porter2005, Roueff2014}). However, the response of ices upon direct vibrational excitation has been scarcely investigated in the laboratory and is typically not included in models (e.g., \citealt{Cuppen2017}). For instance, vacuum ultraviolet radiation is known for resulting in a rich photochemistry (see, e.g., the review by \citealt{Oberg2016} and references therein), transforming the ice structure \citep{Kouchi1990}, as well as leading to photodesorption induced by electronic transitions \citep{Fayolle2011, Fayolle2013, vanHemert2015}, whereas laboratory investigations of potential photodesorption and structural changes upon direct vibrational excitation are still limited. Structural changes are typically reflected in the spectral appearance of such ices (e.g., \citealt{Cuppen2011, Cooke2016, Muller2021}), and therefore changes in ice morphology and content due to energy-dissipation processes will play a part in interpreting observational data from infrared facilities---such as the James Webb Space Telescope \citep{McClure2023}.

Silicates and carbonaceous materials are the main components of (sub)micron-sized dust grains present in the interstellar medium (ISM) \citep{Millar1993}. During the earlier stages of star formation, \ce{H} and \ce{O} atoms adsorbed onto these grains react to form \ce{H2O}, resulting in a water-rich ``polar-ice'' layer \citep{Ioppolo2010, Cuppen2010, Linnartz2015}. Besides water, carbon dioxide (\ce{CO2}) is a very abundant solid-phase ISM species, and is mainly present in the aforementioned interstellar polar-ice layer mixed with \ce{H2O} \citep{Goumans2008, Oba2010, Ioppolo2011, Noble2011}. As the density of the collapsing cloud increases, carbon monoxide (\ce{CO}) molecules present in the gas phase (catastrophically) freeze-out on top of the polar ice, forming a second coating known as the ``apolar-ice'' layer \citep{Tielens1991, Boogert2002, Pontoppidan2003}. Through hydrogenation reactions of solid-state \ce{CO}, methanol (\ce{CH3OH}) is efficiently formed \citep{Tielens1982, Charnley1992, Hiraoka1994, Watanabe2002, Fuchs2009, Santos2022}. This alcohol is abundantly present in interstellar ices mixed with \ce{CO}, as shown by various laboratory and observational works (e.g., \citealt{Bottinelli2010, Cuppen2011, Penteado2015}). Follow-up studies have shown that along the same hydrogenation scheme also larger complex organic molecules (COMs), like glycerol, can be formed involving radical reactions \citep{Chuang2016, Fedoseev2017}. Alternatively, \ce{CH3OH} can also be formed less efficiently during stages prior to \ce{CO}-freeze out through reactions involving \ce{CH4} and \ce{OH} \citep{Qasim2018}.

The infrared irradiation of crystalline water ice has been previously studied with a focus on its resonant desorption at elevated temperatures ($\gtrsim$100 K) \citep{Krasnopoler1998, Focsa2003}. These first studies suggested that on-resonance irradiation of crystalline ices leads to selectivity effects, with a strong wavelength-dependent \ce{H2O} desorption as a result of IR-photon absorption. Infrared-induced matrix assisted laser desorption and ionization (MALDI) experiments ensued, evincing the photodesorption of polycyclic aromatic hydrocarbons (PAHs)---as well as their hydrated clusters---embedded in water ice matrices and subjected to resonant vibrational excitation \citep{Focsa2006, Mihesan2006, Henderson2014}. More recently, porous amorphous solid water (pASW) has been studied in works combining IR-free electron laser (FEL) irradiation experiments and molecular dynamics simulations, which revealed a reorganization of the ice structure towards a more ordered configuration upon the absorption of photons in the broad IR regime \citep{Noble2020, Coussan2022, Cuppen2022}. Amorphous \ce{CO2} ices have shown a similar behavior as a result of IR-FEL irradiation, whereas this effect was barely observed in the crystalline counterparts \citep{Ioppolo2022}. Moreover, both \cite{Noble2020} and \cite{Ioppolo2022} hint at possible photodesorption upon resonant irradiation of the pure ices. 
 
Vibrational energy pooling (VEP) is also an interesting phenomenon that takes place as a result of vibrational excitation. In VEP, vibrationally excited \ce{CO} molecules pool their energy through a base-camp mechanism, resulting in highly excited \ce{CO} species that constitute a small portion of the group \citep{DeLeon1986, Corcelli2002, Chen2017, Chen2019, DeVine2022}. \cite{DeVine2022} have explored the effects of VEP of crystalline \ce{CO} under interstellar conditions in a combined laboratory and theoretical work, and reported the formation of \ce{CO2} and \ce{C2O3} species as a result of spin-forbidden reactions involving highly vibrationally excited \ce{CO} molecules. The reaction pathways they explored also led to \ce{CO} desorption due to the energy release from the dissociation of an intermediate \ce{CO} dimer species only accessible at extremely high vibrational excitation levels. 

Infrared spectroscopic properties such as band width and peak position can provide powerful information on the mixing and layering conditions of ices \citep{Boogert2015}. With the James Webb Space Telescope (JWST), infrared observations of interstellar ices are available at a wide spectral range ($0.6-28.3$ $\mu$m) and with unprecedented spatial resolution, and sensitivity, enabling to focus on specific areas in a proto-planetary disk or to investigate dense interstellar clouds for extinction values as high as 90 \citep{McClure2023}. The interpretation of these observations requires highly detailed laboratory data on the spectroscopic properties, morphology, and dynamics of ices. Moreover, understanding the effects of vibrational energy on the structure and dynamics of the \ce{CO}-rich ice layer plays a key role in developing thorough models of the chemical evolution of the interstellar medium.

In the present work, we explore the effects of IR-photon irradiation and the consequences of vibrational-energy release in interstellar apolar ice analogues. We provide the first selective IR-FEL irradiation study on pure amorphous and crystalline \ce{CO} ices (henceforth \ce{aCO} and \ce{cCO}), as well as pure amorphous \ce{CH3OH} ices and mixtures of the two species. In section \ref{sec:methods}, the experimental setup and procedures employed in this work are detailed. The main results are described and discussed in section \ref{sec:results_diss}. In section \ref{sec:astro} we explore the astrophysical implications, and our main findings are summarized in section \ref{sec:conc}.

\section{Experimental Methods}\label{sec:methods}

\begin{table*}[hbt!]
\centering
\caption{Overview of the experiments performed in this work.}
\label{table:exp_list}      
{\begin{tabular}{cccccc}  
\toprule\midrule
Experiment                  &   T$_{\text{deposition}}$ $^a$   &   T$_{\text{irradiation}}$ $^a$  &   N(\ce{CO}) $^b$              &   N(\ce{CH3OH}) $^b$           &   FEL irradiation\\
                            &   (K)                         &   (K)                         &   (molecules cm$^{-2}$)       &   (molecules cm$^{-2}$)       &   ($\mu$m)\\ 
\midrule
\ce{CH3OH}                  &   20                          &   20                          &                               &   $6\times10^{17}$            & 2.96, 3.07, 3.14, 4.67, 9.62\\
\ce{aCO}                    &   20                          &   20                          &   $1\times10^{17}$            &                               & 4.67, 5.20\\
\ce{cCO}                    &   29                          &   20                          &   $1\times10^{17}$            &                               & 4.67, 5.20\\
\ce{CO}:\ce{CH3OH} = 1:0.3  &   20                          &   20                          &   $2\times10^{17}$            &   $6\times10^{16}$            & 3.07, 4.67, 9.62\\
\ce{CO}:\ce{CH3OH} = 1:3.0    &   20                          &   20                          &   $2\times10^{17}$            &   $5\times10^{17}$            & 3.07, 4.67, 9.62\\
\midrule\bottomrule
\multicolumn{6}{l}{\footnotesize{$^a$ Measured temperature values varied by $\pm$0.8 K.}}\\
\multicolumn{6}{l}{\footnotesize{$^b$ Upper limits derived using transmission band strengths.}}
\end{tabular}}
\end{table*}

The experiments are performed at the HFML-FELIX laboratory at Radboud University, the Netherlands, directing tunable free electron laser radiation towards the laboratory ice surface astrophysics (LISA) ultrahigh vacuum (UHV) end station. At HFML-FELIX, LISA is connected to FELIX-2 and -1 to perform selective irradiation experiments in the broad infrared ($3700-220$ cm$^{-1}$, $2.7-45$ $\mu$m) and terahertz (THz, $10-2$ THz, $30-150$ $\mu$m) spectral ranges respectively, shedding light on the relaxation mechanisms that follow the excitation of particular vibrational modes within solid matter. LISA has been described in detail elsewhere \citep{Noble2020, Ioppolo2022}, and here we provide a brief explanation of the experimental procedure. The experiments described here have been obtained during four 7.5-hours long beam shifts, focusing on the basic concept of (off/on) resonant IR photodesorption of interstellar ices. Extended measurements are scheduled for later this year in another run of 4 beam shifts. 

A gold-plated optically-flat copper substrate in the center of the UHV chamber is in thermal contact with the head of a closed-cycle helium cryostat, which allows to control the substrate temperature in the range of $20-300$ K through resistive heating. The temperature is monitored by a silicon diode thermal sensor fixed at the bottom of the substrate. At room temperature, the base pressure in the chamber is $\sim2\times10^{-9}$ mbar. Gaseous \ce{CO} (Hoekloos/Praxair, purity 99.9\%) and \ce{CH3OH} (Sigma-Aldrich, purity 99.9\%) vapor are introduced in the dosing line, either separately---in the case of pure ices---or simultaneously for a binary ice. Beforehand, the methanol sample is purified through multiple freeze-pump-thaw cycles. The ices are then grown on the substrate through background deposition by admitting the pure species or mixtures into the main chamber via an all-metal leak valve that faces the walls of the chamber. This deposition method leads to more uniform ices and better simulates the deposition conditions on interstellar dust grains. Amorphous \ce{CO} ice is deposited at the lowest substrate temperature of 20 K, while crystalline \ce{CO} is grown at 29 K (see, e.g., \citealt{Kouchi1990b}). The different solid-state configurations are confirmed by the clearly sharper CO-stretching feature of the \ce{cCO} IR spectra. After deposition, all ices are maintained at 20 K during the irradiations.

The ice growth and the effects of the IR irradiation are monitored \textit{in situ} through reflection-absorption infrared spectroscopy (RAIRS) using a Fourier Transform IR spectrometer. The FTIR spectra cover a range of $5000-600$ cm$^{-1}$ with a resolution of $0.5$ cm$^{-1}$. To calculate the column density $(N_X)$ of the species in the ice, a modified Beer-Lambert law is applied to convert the IR integrated absorbance ($\int Abs(\nu)d\nu$) to the absolute abundance:

\begin{equation}
    N_X=\ln10\frac{\int Abs(\nu)d\nu}{A'_X}
\end{equation}

\noindent where $A'_X$ is the absorption band strength of a given species. Since the band strengths of \ce{CO} and \ce{CH3OH} have not been estimated for the LISA setup in reflection mode, we use the transmission values from the literature instead. Namely, $A(\ce{CO})_{\nu_1=2142}=1.1\times10^{-17}$ cm molecule$^{-1}$ \citep{Jiang1975} and $A(\ce{CH3OH})_{\nu_8=1032}=1.8\times10^{-17}$ cm molecule$^{-1}$ \citep{Hudgins1993}. Since the RAIRS technique is more sensitive than transmission infrared spectroscopy, the column densities derived here should be regarded as upper limits. The details of each experiment performed in this work are listed in Table \ref{table:exp_list}.

For the IR irradiation of the ice, we use the IR-FEL source FELIX-2 to generate nearly monochromatic mid-IR photons in the ranges of $2.9 - 4.5$ $\mu$m (fundamental mode) and $4.5 - 9.7$ $\mu$m (third harmonic mode). The photons are bunched in $5-10$ microseconds-long macropulses with a repetition rate of 5 Hz, consisting of a train of micropulses of several picoseconds spaced by 1 nanosecond. Unless otherwise specified, at longer wavelengths ($\gtrsim$4.5 $\mu$m), the energy of the macropulses is attenuated to a constant value of $\sim$20 mJ, as measured at the diagnostic station located before the LISA end-station area of the facility. At shorter wavelengths ($\lesssim$4.5 $\mu$m), the laser energy is kept at its highest attainable value of $\sim$5 mJ, as it is generated in the third harmonic mode. The fluence of infrared photons with wavelength $\lambda$ in each experiment can be calculated by the expression:

\begin{equation}
    \text{Fluence}_\lambda=\frac{E_\lambda*r*t}{S}
\end{equation}

\noindent where $E_\lambda$ is the energy and $r$ is the repetition rate of the macropulses, $S$ is the area of the FELIX-2 beam spot on the substrate, and $t$ is the duration of the irradiation. The FEL beam impinges the gold-plated flat substrate at an angle of $45^{\circ}$ with respect to the surface, resulting in an elliptical spot size of $\sim$1.0 x $\sim$1.5 mm in semi-axes. Comparatively, the FTIR beam angle with the gold substrate is of $13^{\circ}$, resulting in an elliptical spot size of $\sim$1.5 x $\sim$6.5 mm in semi-axes. Consequently, the FTIR RAIRS beam fully covers the region irradiated by the FEL beam, with a FTIR/FEL area ratio of $\gtrsim$6. The spectral FWHM of FELIX-2 is estimated to be $\sim$0.8\% $\delta\lambda/\lambda$ for the entire wavelength range. 

A z-translator allows the movement of the substrate in height, providing multiple ice spots for systematic IR irradiation. In this work we utilize seven different substrate positions, each separated by 4 mm from the neighboring spot. Each FEL irradiation is thereby performed at an unirradiated ice spot, unless specified. The irradiations are subsequently repeated, on average two additional times for each spot, to ensure reproducible results---with the exception of the long \ce{aCO} and \ce{cCO} experiments. As stated above, the FTIR beam is $\gtrsim$6x larger in area than the FEL beam, and thus part of the ice probed by the FTIR is not exposed to FEL irradiation. Hence, difference FTIR spectra acquired before and after FEL irradiation are utilized to visualize changes in the ice. During the irradiation experiments, possible desorption of species from  the ice is monitored by a quadrupole mass spectrometer (QMS). Control temperature-programmed desorption (TPD) experiments with analogous deposition conditions and a ramping rate of 2.5 K min$^{-1}$ are performed to assist in interpreting the IR-FEL irradiation results.

In this paper, we discuss FEL irradiations in terms of wavelength and FTIR spectra in wavenumbers (both in vacuum) to reflect the higher spectral resolution of the FTIR data as opposed to the transform-limited bandwidth of the FEL photons.

\section{Results and Discussion}\label{sec:results_diss}

\begin{figure*}[htb!]\centering
\includegraphics[scale=0.7]{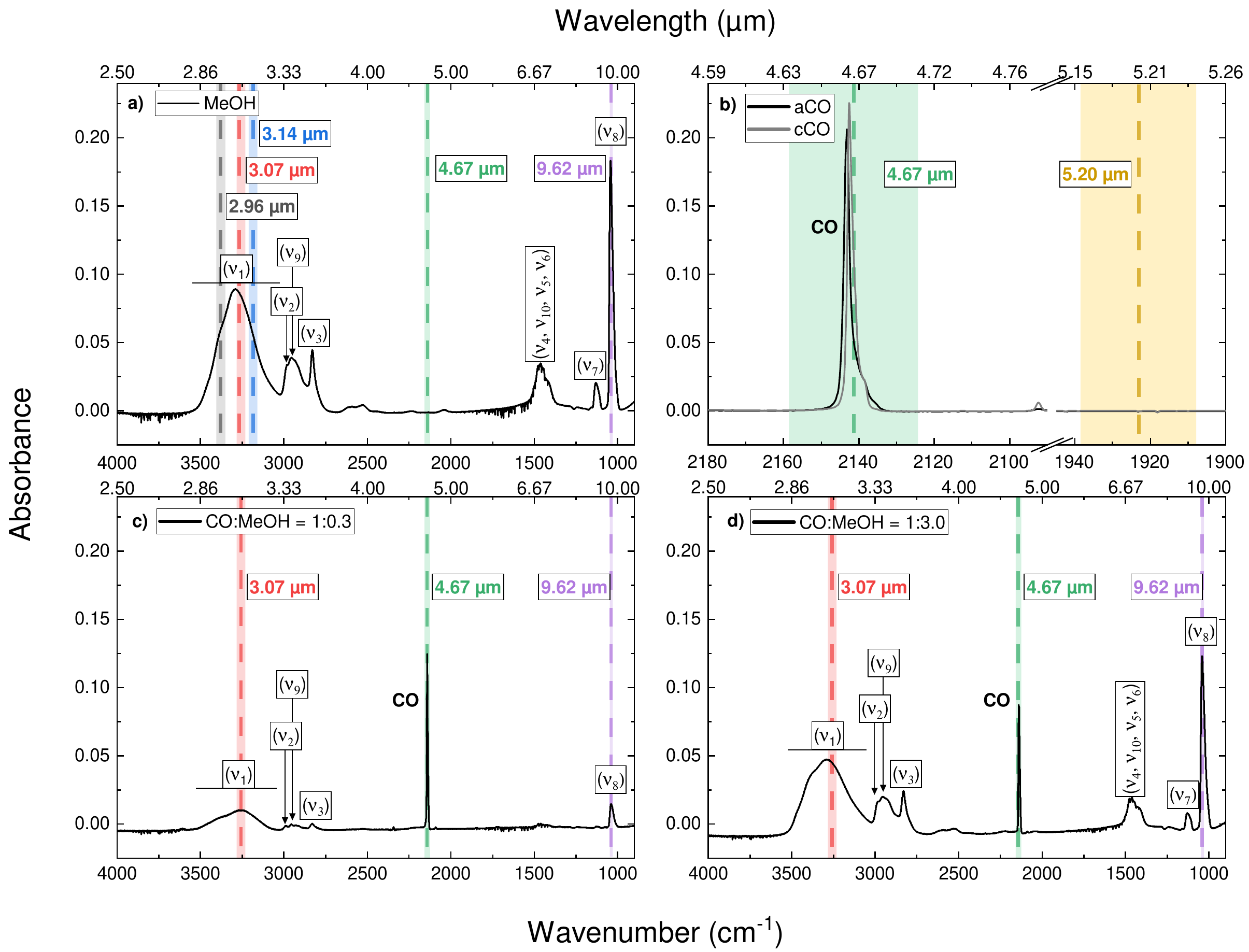}
\caption{Spectra and irradiation maps of the ices explored in this work, all taken at 20 K. Panel a) Pure amorphous \ce{CH3OH}. Panel b) Pure amorphous and crystalline \ce{CO}. Panel c) Mixture of \ce{CH3OH} and \ce{CO} with a \ce{CO}:\ce{CH3OH} ratio of 1:0.3. Panel d) Same as c), but with a 1:3.0 ratio. All spectra are taken after deposition and before irradiation. The wavelengths of FELIX-2 used later for on and off-resonance irradiation are shown with dashed lines, and the FWHM is visible from the shadowed areas. The fundamental mode of \ce{CO} is denoted in boldface, and the other labels correspond to the ($\nu_\text{X}$) vibrational modes of \ce{CH3OH}.}
\label{fig:dep_all}
\end{figure*}

Figure \ref{fig:dep_all} shows the spectra of the ices of pure \ce{CH3OH} (panel a), pure \ce{CO} (both \ce{aCO} and \ce{cCO}, panel b), and 1:0.3 and 1:3.0 mixtures of \ce{CO}:\ce{CH3OH} in panels c and d, respectively. Also indicated in this figure are the involved vibrational modes. The wavelengths of FELIX-2 used during the irradiations are marked by the dashed vertical lines, and their respective FWHMs are shown by the shadowed areas. Given the broad wavelength coverage of the FTIR measurements, the laser bandwidths are hard to see in panels a, c, and d, but the shadowed area is substantially larger in panel b because of the smaller spectral range shown here. For each experiment (i.e., pure \ce{CO}, pure \ce{CH3OH}, and mixtures), we use the same color scheme throughout the paper. Below the individual ices are discussed. 

\subsection{\ce{CH3OH}}\label{subsec:ch3oh}

\subsubsection{Morphology}\label{subsubsec:ch3oh_morph}

\begin{figure*}[htb!]\centering
\includegraphics[scale=0.5]{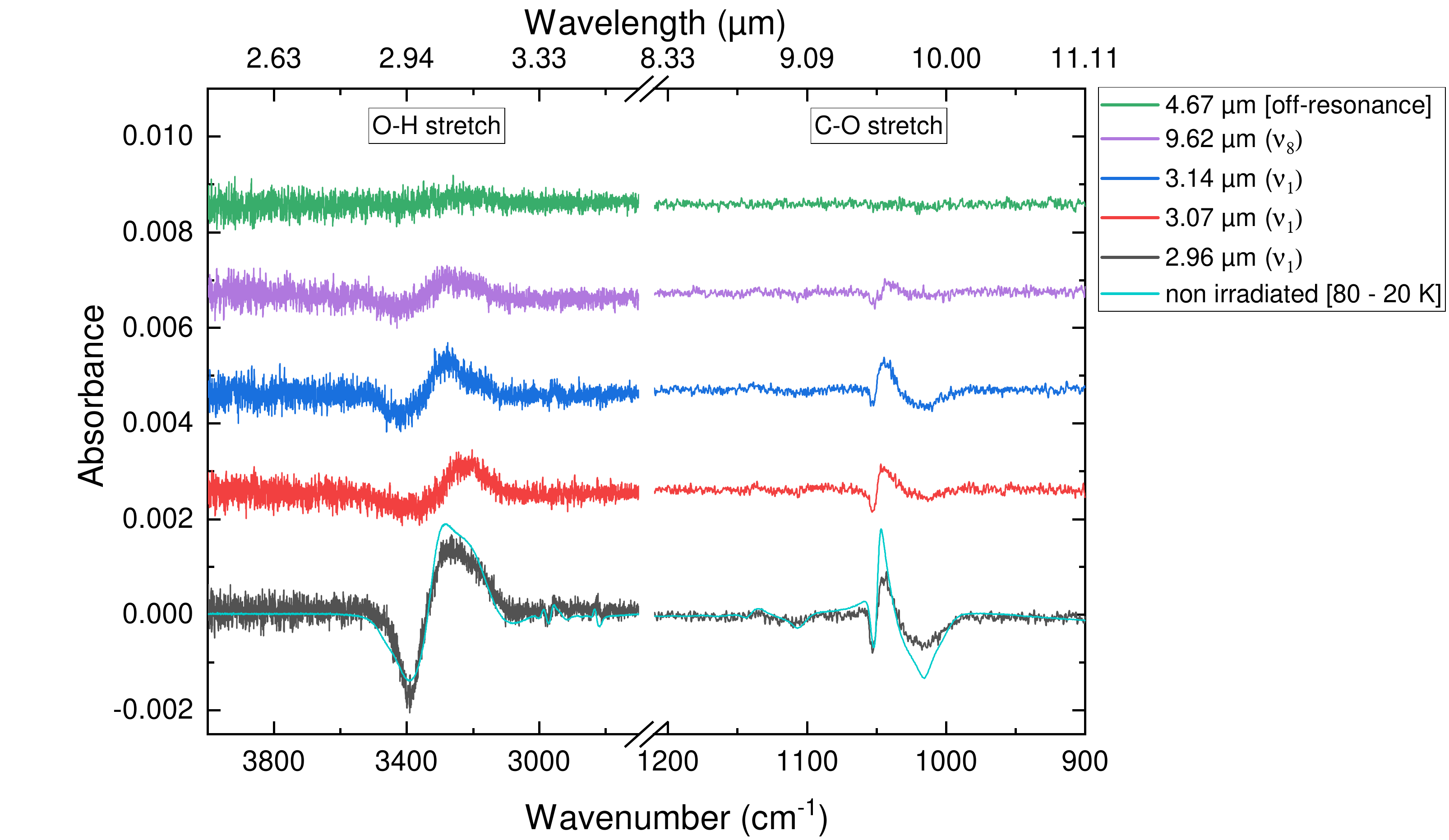}
\caption{Difference spectra obtained before and after 5-minutes IR-FEL irradiation on methanol ice at 20 K. The difference between spectra acquired at 80 K and 20 K (without IR irradiation) during a TPD control experiment is also shown in cyan and included in the 2.96 um plot. To allow direct comparison, all spectra are plotted for the same scale; the differences in the noise level between 3 $\mu$m and 9 $\mu$m are due to the intrinsic wavelength-dependent S/N of the FTIR. The irradiation spectra are offset for clarity.}
\label{fig:meoh}
\end{figure*}

In Figure \ref{fig:meoh}, the difference spectra obtained before and after continuous exposure for 5 minutes of pure \ce{CH3OH} ice to the FEL light at 2.96 ($\nu_1$), 3.07 ($\nu_1$), 3.14 ($\nu_1$), 4.67 (off-resonance), and 9.62 ($\nu_8$) $\mu$m are shown. Intermolecular interactions give rise to the OH-stretching features of \ce{CH3OH} at 3.07 and 3.14 $\mu$m (see, e.g., \citealt{Jakobsen2018, Hudgins1993}), while the shoulder feature at 2.96 $\mu$m has been suggested to originate from molecules with weakly bound O-H bonds, or potentially dangling bonds, due to pores in the ice \citep{Luna2018}. The macropulse energies of all irradiations are adjusted to a constant value of $\sim$5 mJ. Modifications to the \ce{CH3OH} band profiles are clearly observed for all exposures in which the IR wavelength is on resonance with absorption bands of methanol. Comparatively, no differences are observed after the off-resonance irradiation (i.e., at 4.67 $\mu$m). Thus, the observed spectral changes are a consequence of the absorption of IR radiation by the methanol ice, and not by the golden substrate underneath---in agreement with what has been previously observed for pASW and \ce{CO2} ices \citep{Noble2020, Ioppolo2022}. Overall, each \ce{CH3OH} absorption feature shows similar spectral changes in the difference spectra of all on-resonance irradiations, with only variations in their intensities. The most pronounced effects are observed for the irradiations at the OH-stretching region. Indeed, one may expect that irradiations at the OH-stretch---which has a higher band strength---would result in more photon-absorption events than those on resonance with the CO-stretching mode. Furthermore, the exposure at the \ce{CH3OH} $\nu_1$ band at 2.96 $\mu$m, in particular, yields changes significantly stronger than the other OH stretches. The fact that neighboring irradiations in the OH-stretching region yield such different results highlights the high dependence of the energy-dissipation mechanisms on the type of vibrational mode excited, even within similar energy ranges. Irradiations at the CH-stretching modes of \ce{CH3OH} are also performed (not shown) and result in similar spectral changes to the other on-resonance spectra, albeit with lower intensities.

The features around the OH- and CO-stretching bands of methanol in the difference spectra show both positive and negative components that yield a negligible net integral absorbance. Those features are thus attributed to the restructuring of the ice upon exposure to the IR-FEL with a power of $\sim5$ mJ, leading to an overall loss of the weakly bound O-H modes at 2.96 $\mu$m and corresponding gain of bulk O-H modes at 3.07 and 3.14 $\mu$m. A similar behavior was observed for O-H dangling modes versus bulk modes of IR-FEL irradiated pASW \citep{Coussan2022}. Moreover, the CO-stretching band profiles show an overall narrowing trend, typically of the order of $0.2\%$. It should be noted that the significantly larger area of the FTIR beam in comparison with the IR-FEL beam results in an appreciable ($\gtrsim$6x) dilution of the laser effect on the recorded IR band shapes, meaning that the aforementioned percentage is in fact a lower limit.

To investigate the character of the morphological changes induced by the IR-FEL, we performed a control TPD experiment of unirradiated pure \ce{CH3OH} ice deposited under the same conditions as in the irradiation study. Such an experiment allows to compare controlled thermally induced morphology changes with those induced upon IR irradiation. The IR spectra acquired during the TPD experiments are reported in the Appendix \ref{appendix:control_IR}. The spectra as obtained for different temperatures were compared to the difference spectra before and after irradiation. The best qualitative fit to the most effective irradiation at $\nu_1$---i.e., the TPD result that best reproduces the irradiation---is obtained by the difference between the spectra taken at 80 K and 20 K, as shown in Figure \ref{fig:meoh}. The profile of the IR spectrum obtained during the TPD experiment reproduces quite well that of the irradiation spectrum for the OH-stretching modes, whereas discrepancies arise in the intensity of the changes for the CO-stretching counterpart. Nonetheless, the overall profiles of both difference spectra acquired during irradiation and the TPD experiment are remarkably similar. This is a strong indication that, for pure methanol ices, the excess vibrational energy is efficiently dissipated throughout the bulk of the ice in an equivalent manner to thermal heating of an extended area. This energy dissipation leads to restructuring towards a more organized morphology, as was concluded for pASW and \ce{CO2} ices \citep{Noble2020, Ioppolo2022, Coussan2022}.

The polarity of the ice could play a role in the effectiveness of the vibrational energy dissipation through restructuring: similarly to the case of methanol ice reported here, \cite{Noble2020} have observed that the difference spectra upon IR-FEL irradiation of pASW can be satisfactorily reproduced by the subtraction of theoretical spectra simulating a restructuring of the ice. However, \cite{Ioppolo2022} found that for \ce{CO2} ices local restructuring alone is not sufficient to explain the changes induced by the IR-FEL. Alternatively, this difference could also be related to the type of intermolecular interactions within the ice, in which \ce{H}-bonding networks could potentially facilitate the vibrational energy dissipation through restructuring, while generally weaker van der Waals interactions could be less efficient in doing so. Indeed, transfer of vibrational energy has been shown to occur in \ce{H2O} ices through hydrogen-bonded water molecules with resonant O-H stretches---which leads to local heating and restructuring of the ice \citep{Cuppen2022}.

\subsubsection{Power-dependence analysis and photodesorption}\label{subsubsec:ch3oh_photodes}

\begin{figure}[htb!]\centering
\includegraphics[scale=0.4]{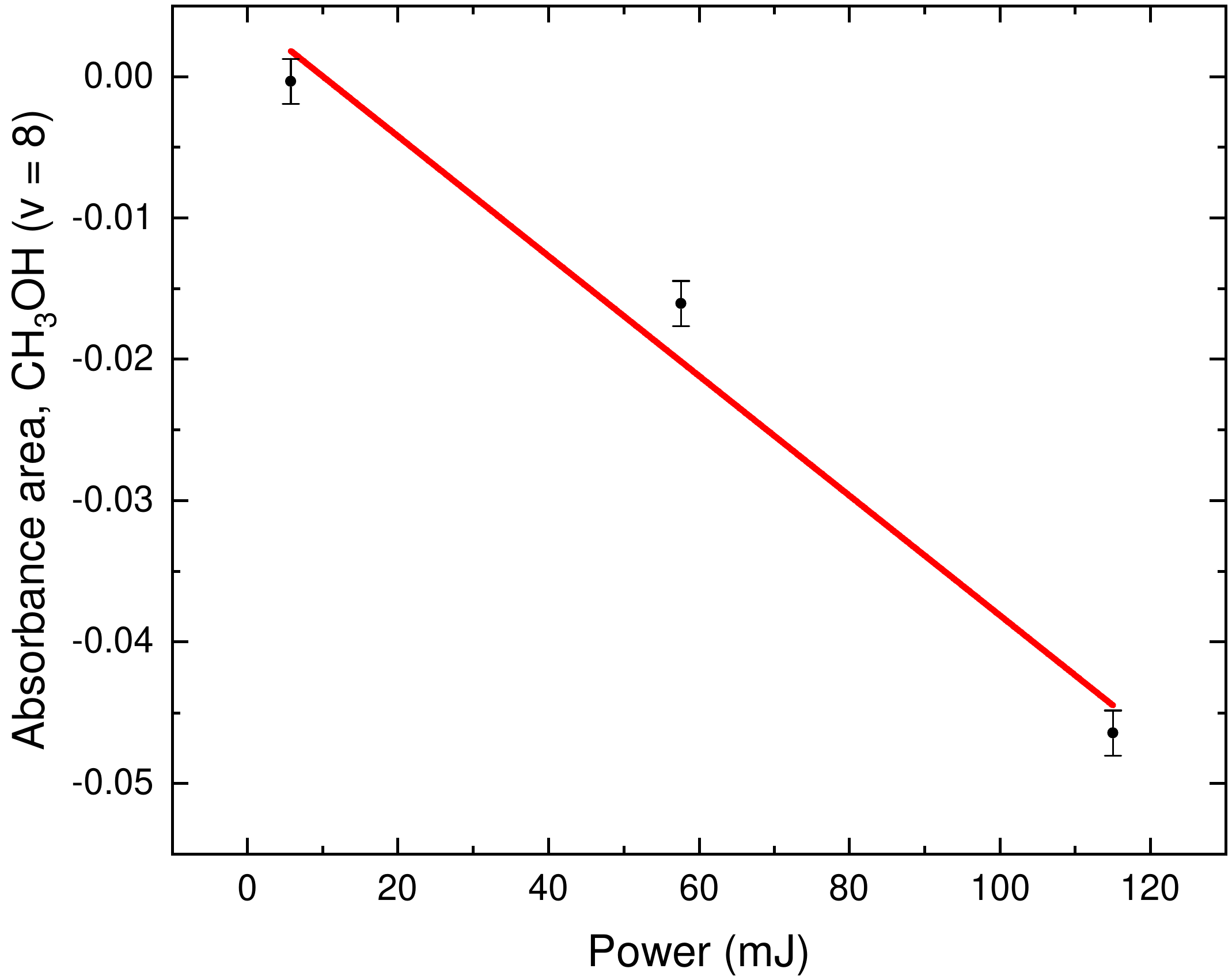}
\caption{Power-dependence analysis of the IR-FEL effect on \ce{CH3OH} ice. The three different power settings---5.76, 57.6, and 115 mJ---are controlled by attenuating a 115 mJ beam with 13, 3, and 0 dB, respectively. Error bars correspond to the three sigma limit of the instrumental error as derived from the integrated noise signal of the off-resonance difference spectrum in the same band width.}
\label{fig:power}
\end{figure}

It is possible to assess whether the changes to the \ce{CH3OH} ice are due to single- or multi-photon processes through a power-dependence analysis. We performed IR-FEL irradiations of a pure \ce{CH3OH} ice sample at 9.62 $\mu$m with three different power settings (5.76 mJ, 57.6 mJ, and 115 mJ) for 5 minutes and compared the resulting integrated net areas around the $\nu_8$ absorption band of methanol as a function of IR-FEL power. Figure \ref{fig:power} shows a fairly linear (R$^2= 0.98$) correlation between beam power and area, further suggesting that the effect of the IR-FEL laser on the ice morphology is due to single-photon processes \citep{Noble2020, Ioppolo2022, Coussan2022}. The decreasing trend in the integrated absorption signal at the $\nu_8$ band with increasing beam power suggests that photodesorption might be observed in the infrared spectra once enough vibrational excitations take place in the ice. The extent of the desorption is however not high enough to be detectable with the sensitivity of the current QMS. From the integrated absorption signal of the \ce{CH3OH} $\nu_8$ band lost after the 115 mJ irradiation and the total beam fluence---$\sim1.8\times10^{23}$ photons cm$^{-2}$---we derive a tentative upper limit to the photodesorption rate of $r=(1.6\pm0.5)\times10^{12}$ molecules J$^{-1}$, or $r=(3\pm1)\times10^{-8}$ molecules photon$^{-1}$. We consider only the uncertainties in the band strength to derive the errors in the photodesorption rates, although other error sources could also have an influence. The restructuring of the ice is expected to have a minor effect on the band strength of \ce{CH3OH}, as it was shown to only change appreciably at temperatures above the crystallization of methanol \citep{Luna2018}.

The resulting rate is orders of magnitude lower than typical (non-dissociative) \ce{CH3OH} UV photodesorption counterparts, of $10^{-5}$ molecules photon$^{-1}$ \citep{Bertin2016}, but it is still measurable because of the much higher IR flux of FELIX-2 compared to the UV-broadband flux of a microwave \ce{H2} discharge lamp \citep{Ligterink2015, Cruz-Diaz2016} or monochromatic radiation of a synchrotron VUV beam \citep{Fayolle2011}. At a first glance, the astronomical relevance of such a low photodesorption rate may be considered negligible, but one should realize that in dense interstellar clouds the IR photon fluxes are $\gtrsim10^4$ times higher than that of the secondary UV, resulting in comparable efficiencies for IR-induced photodesorption phenomena in the interstellar medium. These efficiencies and their impacts on astrochemical models are further discussed in section \ref{sec:astro}. Furthermore, whereas UV studies show that \ce{CH3OH} largely fragments upon electronic excitation \citep{Bertin2016}, it is likely that upon IR excitation desorption follows a non-dissociative pathway---further influencing the predicted abundance of \ce{CH3OH} in the gas phase.

\subsection{CO}\label{subsec:co}

\subsubsection{Morphology}\label{subsubsec:co_morph}

\begin{figure*}[htb!]\centering
\includegraphics[scale=0.5]{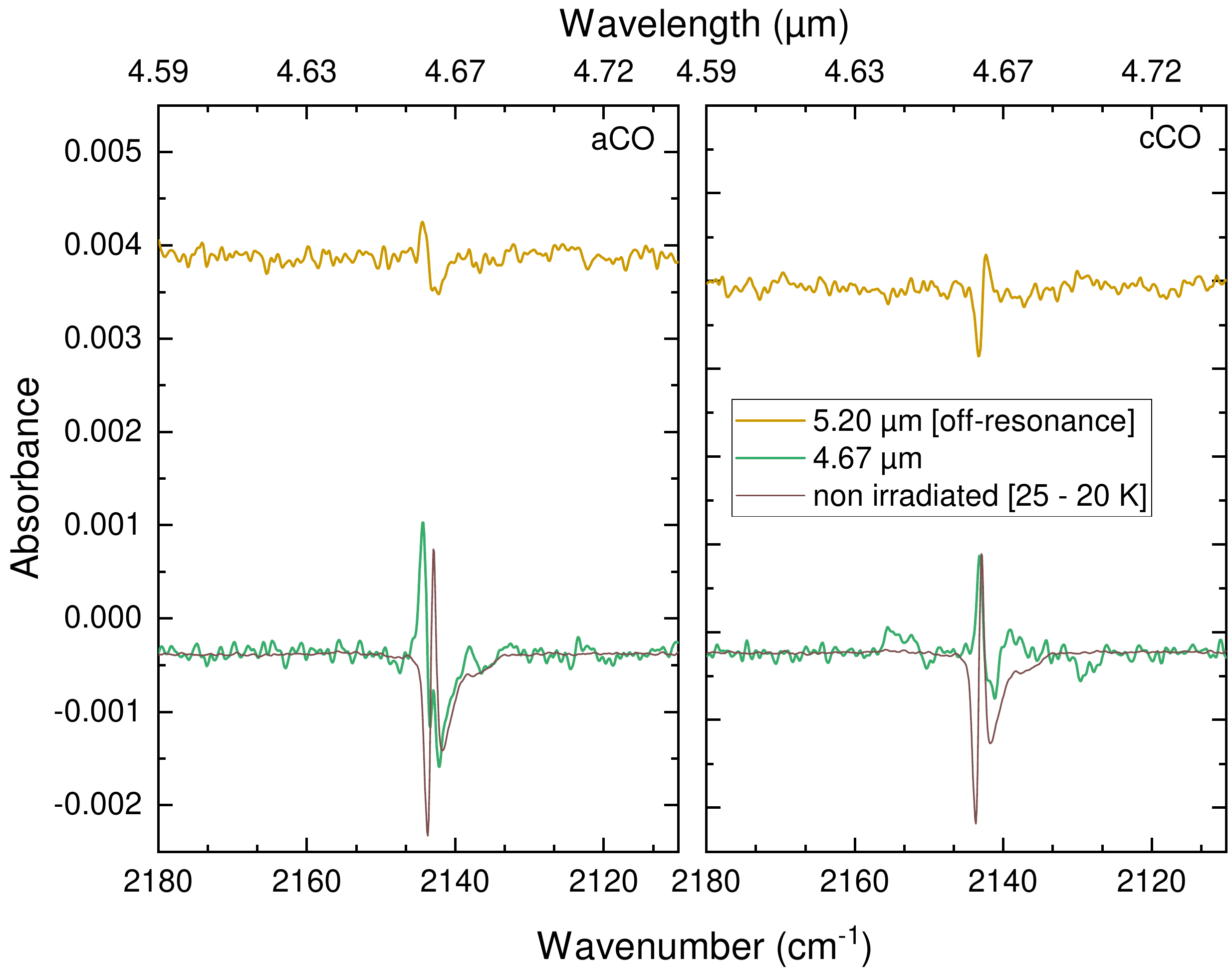}
\caption{Difference spectra obtained before and after 10-minutes IR-FEL irradiation on \ce{aCO} ice (left panel) and \ce{cCO} ice (right panel), both at 20 K. The difference between spectra acquired at 25 K and 20 K during a TPD control experiment is shown in brown. The irradiation spectra are offset for clarity.}
\label{fig:CO_diff}
\end{figure*}

The difference spectra obtained before and after 10-minutes IR-FEL exposure of amorphous and crystalline \ce{CO} ice at 4.67 and 5.2 $\mu$m are shown in Figure \ref{fig:CO_diff}. Since in the CO-ice case there is only one vibrational absorption band available to resonant irradiation---limiting the number of excitations to perform and thus leaving more time to perform longer irradiations---we opted for a higher IR-FEL fluence (i.e., 10 instead of 5 minutes irradiation) in order to further increase the signal-to-noise ratio in the IR difference spectra. Irradiations of \ce{CO} ice at wavelengths off-resonance (i.e., at 5.20 $\mu$m) show a feature that turns out to be due to the ongoing stabilization of the ice---a process that extends to a time frame of several hours after deposition. This is confirmed by recording spectra taken 10 minutes apart (without any exposure to the IR-FEL) directly before performing the irradiations. Their difference spectra shows similar features with identical intensities to the one observed in the 5.20 $\mu$m case. For direct comparison, the infrared spectra of the 4.67 $\mu$m and 5.20 $\mu$m irradiations corrected for the artifacts due to ice stabilization are shown in Appendix \ref{appendix:co_stab}.

Due to the intrinsic band width, the irradiation at 4.67 $\mu$m overlaps with both the longitudinal optical (LO) and transverse optical (TO) phonons of the vibrational mode of CO---at $\sim$2142 and $\sim$2138 cm$^{-1}$, respectively. The on-resonance FEL exposure at 4.67 $\mu$m yields more pronounced ($\sim$2x) modifications to the ice in the amorphous case than in the crystalline counterpart. This is in line with the hypothesis that the ice undergoes some degree of restructuring toward a more organized configuration, in which case \ce{cCO} would reach a saturation point more quickly, as its structure is already largely organized. Similar saturation effects have been reported in previous studies on crystalline \ce{H2O} and \ce{CO2} ices \citep{Noble2020, Ioppolo2022}.

The beam exposure at 4.67 $\mu$m is compared to the TPD difference spectrum between 25 K and 20 K, also shown in both panels in Fig. \ref{fig:CO_diff}. As shown by the difference spectrum from the TPD experiment, heating of the ice causes the \ce{CO} molecules to reorganize in a more compact configuration, resulting in a narrower band profile. Additionally to the narrowing effect, the irradiation spectrum also shows a slight overall blueshift of the \ce{CO} band, suggesting that repulsive interactions between the molecules might become more prominent upon restructuring. Concomitant ice stabilization could contribute to the apparent blueshift, albeit to a smaller extent. Ultimately, the spectral changes induced by the IR-FEl cannot be simply reproduced by the TPD-IR results---in contrast to the case of \ce{CH3OH} shown in Figure \ref{fig:meoh}. This suggests that the dissipation of vibrational energy throughout the \ce{CO} ice is not as efficient as for \ce{CH3OH}, and thus the effects of the IR-FEL exposure in the former case cannot be described solely by an extended heating that transforms the structure of the ice.

\subsubsection{Photodesorption}\label{subsubsec:co_photodes}

\begin{figure}[htb!]\centering
\includegraphics[scale=0.35]{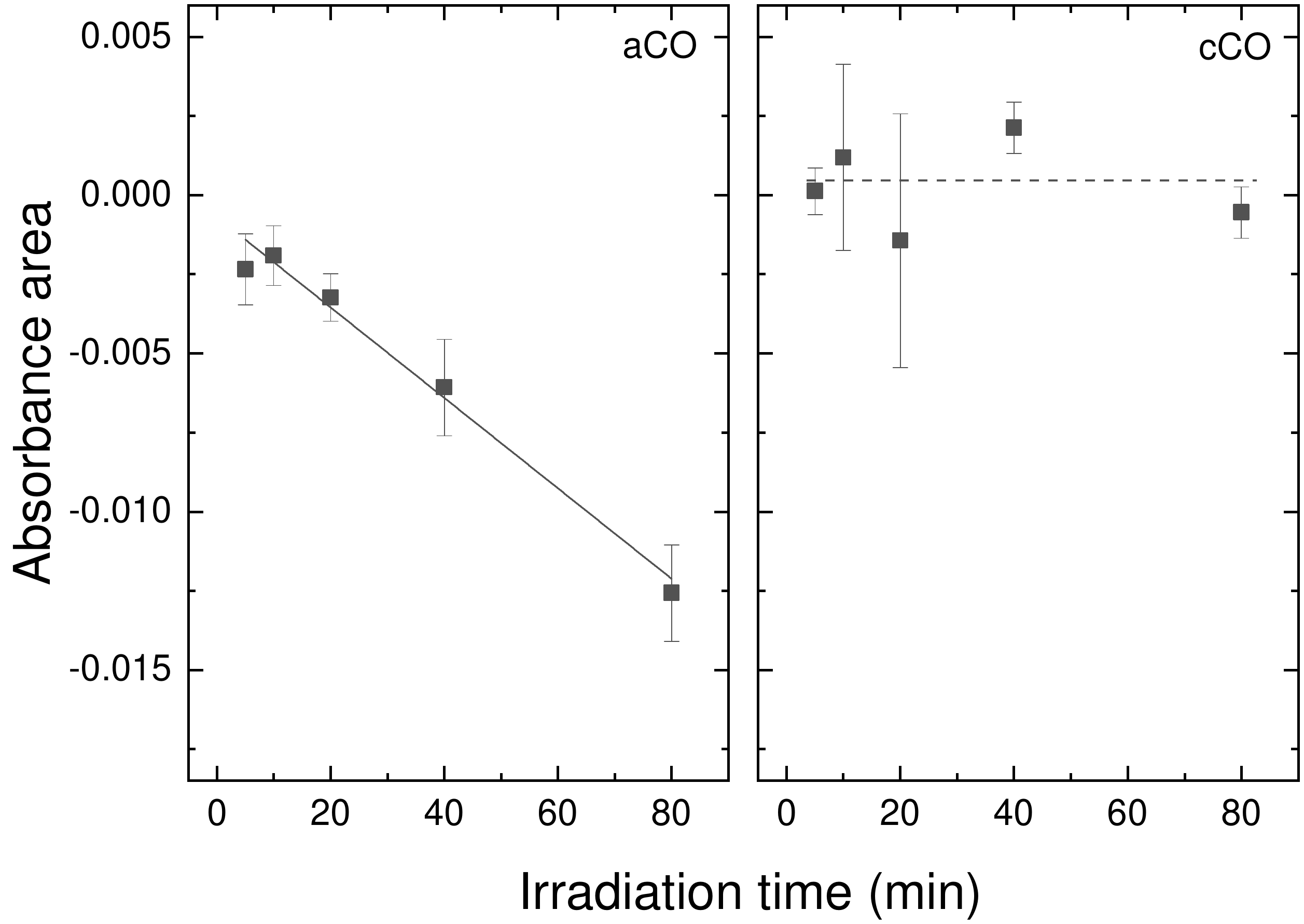}
\caption{Plot of the integrated area of the stretching mode of \ce{CO} in the difference spectra obtained as a function of irradiation time. Left panel: Irradiation of amorphous \ce{CO} ice. The solid line shows the linear fit to the points. Right panel: Irradiation of crystalline \ce{CO} ice. The averaged integrated noise is denoted by the dashed line.}
\label{fig:CO_areas}
\end{figure}

To thoroughly investigate the temporal effects of vibrational energy injection into pure \ce{CO} ice and the possibility of photodesorption, we systematically irradiated both the amorphous and crystalline samples at 4.67 $\mu$m for different irradiation durations. In Figure \ref{fig:CO_areas}, the integrated net area around the \ce{CO} stretching mode obtained from the difference spectra before and after each irradiation are shown as a function of irradiation time (i.e., fluence). The error bars include the fluctuations in integrated absorbance area due to the stabilization of the ice, such as the artifacts seen after the irradiation at 5.20 $\mu$m in Figure \ref{fig:CO_diff}.

The overall absorbance area of the stretching feature of \ce{aCO} decreases linearly with irradiation fluence, strongly suggesting a continuous loss of material upon exposure to the beam due to IR-induced photodesorption. On the contrary, there is no significant loss in the intensity of the stretching mode of crystalline \ce{CO} due to irradiation for the fluences explored in this work. Thus, \ce{cCO} does not seem to photodesorb (as effectively) as a result of IR-photon absorption to the same extent that \ce{aCO} does. This contrasts with results on \ce{H2O} ices irradiated by IR-FEL reported by \cite{Noble2020}, in which case pASW was observed to mainly restructure upon vibrational excitation, with only a minor potential contribution from photodesorption. On the other hand, irradiated cubic crystalline ices underwent pure photodesorption, totaling a loss of $\sim$15\% of the deposited material. The different nature of the intermolecular interactions among \ce{CO} and \ce{H2O} ices could be the reason behind this discrepancy. \cite{vanHemert2015} predicted the UV-induced photodesorption probability of \ce{cCO} to be around 5 times smaller than that of \ce{aCO}, which they attributed to the stronger binding of the molecules in the crystalline structure. This value is remarkably consistent with experimental measurements of \ce{CO} photodesorption as a function of deposition temperature \citep{Oberg2009}. Likewise, the same effect could also contribute to the significantly lower desorption efficiency of \ce{cCO} upon FEL irradiation. Since the observed decrease in the \ce{CO} band intensity does not happen at the precise wavelength of irradiation---which is shifted from the \ce{CO} peak---we can rule out the contribution from hole-burning effects to the negative integrated areas.

The transfer of vibrational energy to translational energy (i.e., vibrationally induced photodesorption) by \ce{CO} molecules has been previously predicted to be inefficient \citep{vanHemert2015, Fredon2021}. In this work, we observe a total absolute loss of $\sim2.6\times10^{15}$ molecules cm$^{-2}$ of \ce{aCO} as a result of 80 minutes of irradiation, which yields a relatively small amount of desorbed species that is below the detection limit of the QMS---rendering the mass spectrometry data inconclusive. Nonetheless, our infrared spectroscopy data provide clear evidence of a \ce{CO} photodesorption process that has not been predicted by the aforementioned models. From the total column density loss of \ce{aCO} after 80 minutes of irradiation and the total beam fluence integrated over the entire exposure time---$\sim2.4\times10^{23}$ photons cm$^{-2}$---we derive a tentative upper limit to the photodesorption rate of $r\sim(2.6\pm0.8)\times10^{11}$ molecules J$^{-1}$, or $\sim(1.1\pm0.3)\times10^{-8}$ molecules photon$^{-1}$. The errors were estimated from the band strength uncertainties. Similarly to \ce{CH3OH}, changes in the band strength of \ce{CO} due to restructuring are expected to be minor, and the derived rate translates to a \ce{CO} desorption efficiency induced by IR photons comparable to the UV value under interstellar conditions (see section \ref{sec:astro}). Moreover, the linear trend between the decrease in integrated absorption signal and photon fluence is further evidence of a zeroth-order, single-photon interaction of the IR beam with the ice leading to desorption---as has been observed before for the UV-induced photodesorption of \ce{CO} \citep{Oberg2007, MunozCaro2010, Fayolle2011, Chen2014, Munoz-Caro2016, Paardekooper2016}. Similar to this work, other studies of FEL exposure to pASW and \ce{CO2} ices suggest the occurrence of IR-induced photodesorption (see, e.g, \citealt{Noble2020, Ioppolo2022, Coussan2022}).

In the case where multiple \ce{CO} species are vibrationally excited simultaneously, effects such as VEP phenomena can arise. In the interstellar medium this is rather unlikely, but for the high fluxes used in the laboratory this should not be a priori neglected. As a consequence, some desorption mechanisms could in principle become accessible to \ce{CO} molecules that collect energy from their neighbors and reach higher vibrational levels \citep{DeVine2022}. Differently from \cite{DeVine2022}, however, we do not detect any photoinduced products upon irradiation of \ce{aCO} nor \ce{cCO}. This is likely due to the differences between the experimental conditions of both studies---e.g., ice morphology, thickness, IR-photon generation method---that result in a significant quenching of the pooling effect during our experiments in comparison to \cite{DeVine2022}. Indeed, simultaneously excited species at any given time are expected to be very diluted within the ice in this work, as the number of absorption events per IR-FEL pulse is estimated to be much lower ($\sim$0.1\%) than the total ice column density. Excited species are therefore expected to be sufficiently apart to prohibit effective interactions.. Moreover, it is worth emphasizing that multi-photon excitations are unlikely under our experimental conditions, as is shown by the linear trend in Figure \ref{fig:power}. 

\subsection{Mixtures}\label{subsec:mixtures}

\subsubsection{Morphology}\label{subsubsec:mix_morph}

\begin{figure*}[htb!]\centering
\includegraphics[scale=0.5]{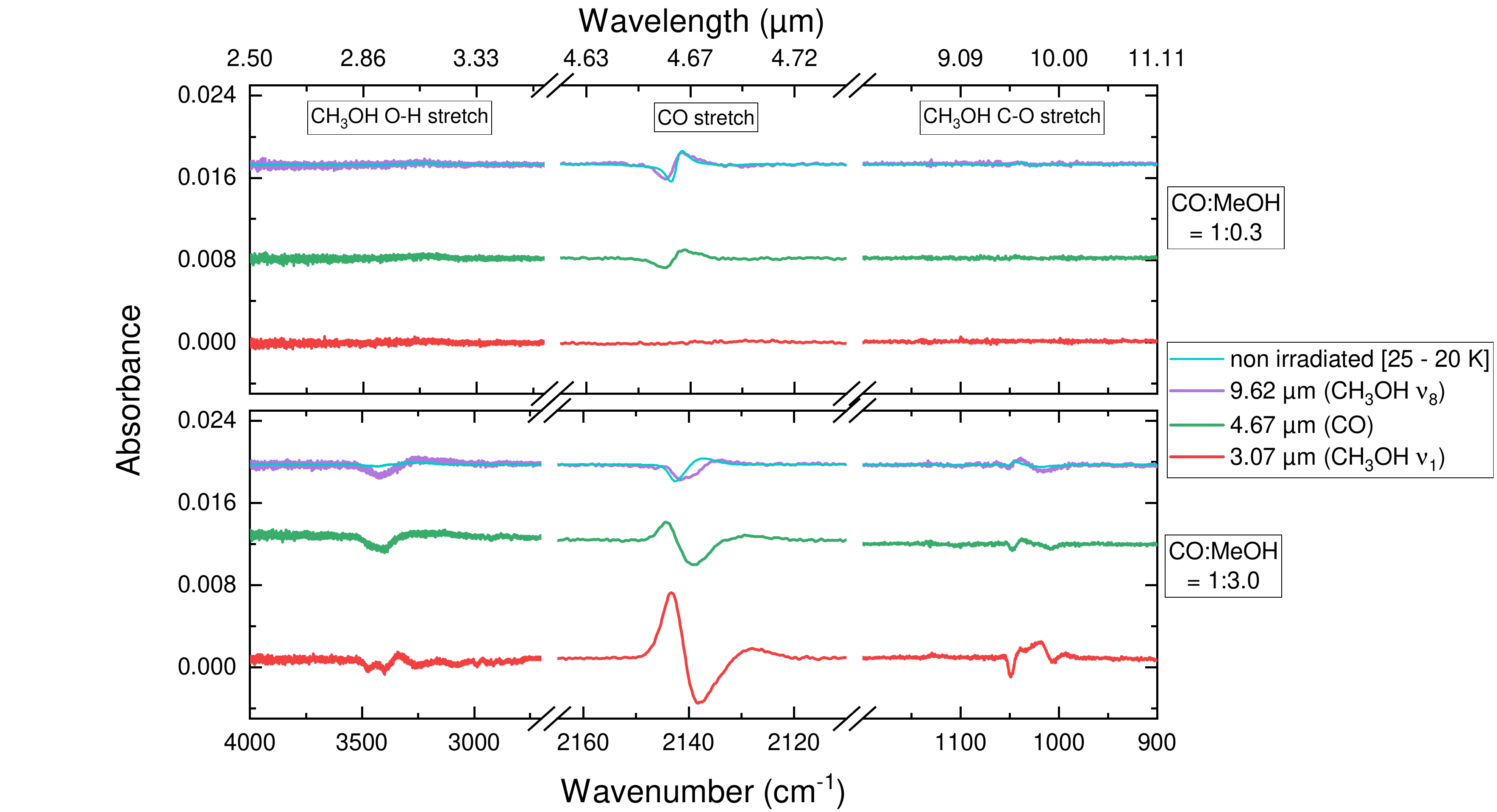}
\caption{Difference spectra obtained before and after 5-minutes IR-FEL irradiation on ice mixtures of \ce{CO} and \ce{CH3OH} at 20 K. Upper panel: Mixture ratio of \ce{CO}:\ce{CH3OH} = 1:0.3. Lower panel: Mixture ratio of \ce{CO}:\ce{CH3OH} = 1:3.0. The control TPD difference spectra between 25 K and 20 K are also shown in cyan. The irradiation spectra are offset for clarity.}
\label{fig:mix_diff}
\end{figure*}

To further explore the impact of the chemical environment on the dissipation of vibrational energy, we performed irradiation experiments on ice mixtures of \ce{CO} and \ce{CH3OH} with ratios of \ce{CO}:\ce{CH3OH} = 1:0.3 (\ce{CO}-rich) and 1:3.0 (\ce{CH3OH}-rich). As methanol is expected to form through CO hydrogenation, such mixtures are of direct astronomical relevance \citep{Watanabe2002, Fuchs2009, Cuppen2011}. The difference spectra of both mixtures obtained before and after 5 minutes of FEL exposure at 20 K are shown in Figure \ref{fig:mix_diff}. For comparison, we maintain the same absolute column densities of \ce{CO} in both CO-rich and methanol-rich ices (see Table \ref{table:exp_list}). 

The FEL exposures of the CO-rich ice do not result in any significant changes around the methanol features (i.e., CO and OH stretches), namely above the detection limit (top panel of Figure \ref{fig:mix_diff}). This is likely because the impact of restructuring changes of the deposited amount of methanol remains within the spectral noise level. The CO feature, conversely, is modified by the IR-FEL irradiation at both 4.67 and 9.62 $\mu$m, displaying the same spectral profile in both cases. After 5 minutes of irradiation with $\sim$20 mJ, the ice becomes saturated to the IR-FEL effects (see Figure \ref{fig:CO_rich_app} in Appendix \ref{appendix:repeated_mix}). This indicates that the absorption of photons by either \ce{CO} or \ce{CH3OH} molecules (at 4.67 and 9.62 $\mu$m, respectively) leads to similar modifications of the ice, which result in a slight redshift of the overall \ce{CO} band profile. The IR difference spectrum obtained during TPD between 25 K and 20 K reproduces the \ce{CO} shape reasonably well (see upper graph in the top panel), meaning that most ice changes can be explained by an extended heating-like event upon photon absorption that leads to a more organized configuration. However, the negative features in the irradiation spectra are not completely reproduced by the TPD control experiment, which suggests that additional desorption could also be taking place. It is worth noting that both the irradiation and TPD difference spectra of the CO-rich mixture display a different profile around the \ce{CO} band compared to the pure \ce{aCO} ice. In the TPD experiments, pure amorphous \ce{CO} becomes narrower with higher temperatures as a result of an increasingly ordered configuration, but no change in peak position is observed. In the binary ice, conversely, the CO stretching mode becomes slightly redshifted at higher temperatures, indicating that (more) attractive interactions dominate the new \ce{CO} sites. These observations highlight the crucial role played by the surrounding species on solid-state IR band shapes. Moreover, the on-resonance irradiation at the $\nu_1$ vibrational mode of \ce{CH3OH} at 3.07 $\mu$m results in no apparent changes around the \ce{CO} feature, despite it yielding the most pronounced effects in the case of pure methanol ice. These observations provide further evidence of the strong dependence of the dissipation of vibrational energy on the involved excited vibrational mode and on interactions with the surrounding ice matrix.

In the case of the methanol-rich ice, modifications to both species' bands upon irradiation are observed above the detection limit of the instrument (bottom panel of Figure \ref{fig:mix_diff}). The higher absolute abundance of methanol in this case compared to the CO-rich ice allows for more IR photons to be absorbed when irradiation is on resonance with methanol absorption wavelengths. As a consequence, the exposures at 9.62 $\mu$m and especially at 3.07 $\mu$m result in significantly more intense changes to the \ce{CO} band than in the \ce{CO}-rich ice. Whereas in the latter case the irradiation at the $\nu_1$ mode of \ce{CH3OH} leads to no differences in the \ce{CO} band, in a more methanol-rich environment---likewise in pure methanol ice---it yields the most pronounced profile changes. The methanol features in this mixture, however, display a remarkably different behavior upon irradiation at its OH stretching mode in comparison to the pure methanol case. Clearly, the changes in the band profiles of both \ce{CH3OH} and \ce{CO} species are highly dependent on the excited vibrational mode, as found by the differences of the shapes resulting from the three irradiation frequencies. Interestingly, when irradiated at 3.07 $\mu$m and 4.67 $\mu$m, the profile of the stretching feature of \ce{CO} in the methanol-rich mixture is the inverse of what is observed for the \ce{CO}-rich counterpart: while the IR excitation of the former leads to a blueshift of the peak, in the latter it results in a redshift. In fact, the blueshifted profile better resembles the one observed in the irradiation of pure \ce{aCO}, despite their very different surrounding conditions. These profiles are repeatedly observed during additional measurements of the \ce{CH3OH}-rich ice mixture (see Figure \ref{fig:CH3OH_rich_app} in Appendix \ref{appendix:repeated_mix}). However, the mechanisms that result in such effects are still unclear. Further systematic investigation of the dissipation of vibrational energy by \ce{CO}:\ce{CH3OH} ice mixtures with various ratios are warranted to fully understand these observations.

Both irradiations at 3.07 and 4.67 $\mu$m lead mainly to a decrease in the area of the \ce{OH}-stretching band of methanol, as well as a narrowing of its \ce{CO} stretching mode. Comparatively, the irradiation at 3.07 $\mu$m results in a more complex difference spectrum, which shows methanol vibrational modes that do not follow a clear trend in overall band shift nor width change. Furthermore, despite both ices containing the same column density of \ce{CO} molecules, the higher abundance of surrounding \ce{CH3OH} species in the methanol-rich ice results in more intense changes to the \ce{CO} band, even upon irradiation at 4.67 $\mu$m. As was observed by \cite{Cuppen2022} in \ce{H2O} ices, it is likely that the higher fraction of hydrogen-bonded species in the \ce{CH3OH}-rich mixture facilitates the vibrational energy transfer in comparison to the \ce{CO}-rich counterpart, thus resulting in more intense spectral changes. In both ice mixtures, irradiations at the CH-stretching bands of methanol (not shown) yield similar changes to the other modes, but less intense. 

\cite{Fredon2021} have explored the dissipation of different types of energy (i.e., vibrational, rotational and translational) by admolecules on top of an ASW surface using molecular dynamics simulations. They observed that the distribution of vibrational energy among the species' vibrational modes heavily impacts the energy dissipation channels, in agreement with our experimental results and those of previous IR-FEL exposure works \citep{Noble2020, Ioppolo2022, Coussan2022, Cuppen2022}. They also concluded that the dissipation of vibrational energy from the admolecule to the surface occurs through the excitation of a surface-admolecule bond, and therefore intermolecular interactions play a key role in this process---which is corroborated by our experimental results. Further theoretical works on the dissipation of vibrational energy in ice mixtures are needed to fully understand the profile changes observed in Figure \ref{fig:mix_diff}.

As for the CO-rich case, the irradiation spectrum of the methanol-rich ice at 9.62 $\mu$m is best reproduced by the corresponding TPD difference spectrum between 25 and 20 K. Yet, neither the band profiles nor the relative intensities can satisfactorily match that of the 9.62 $\mu$m irradiation spectrum, which displays overall more negative features than the TPD control experiment. This indirectly implies that consumption of the ice material should take place, for instance, through photodesorption of \ce{CO}, \ce{CH3OH}, or both. We do not detect any new signals in the IR spectra after irradiation, and thus IR-induced photochemistry is unlikely to contribute to this effect. Additionally, the irradiations at 3.07 $\mu$m and 4.67 $\mu$m yield difference spectra that cannot be reproduced by any combination of the TPD control spectra, and thus must be dominated by a process other than extended heating.

\subsubsection{Photodesorpion}\label{subsubsec:mix_photodes}

\begin{figure*}[htb!]\centering
\includegraphics[scale=0.6]{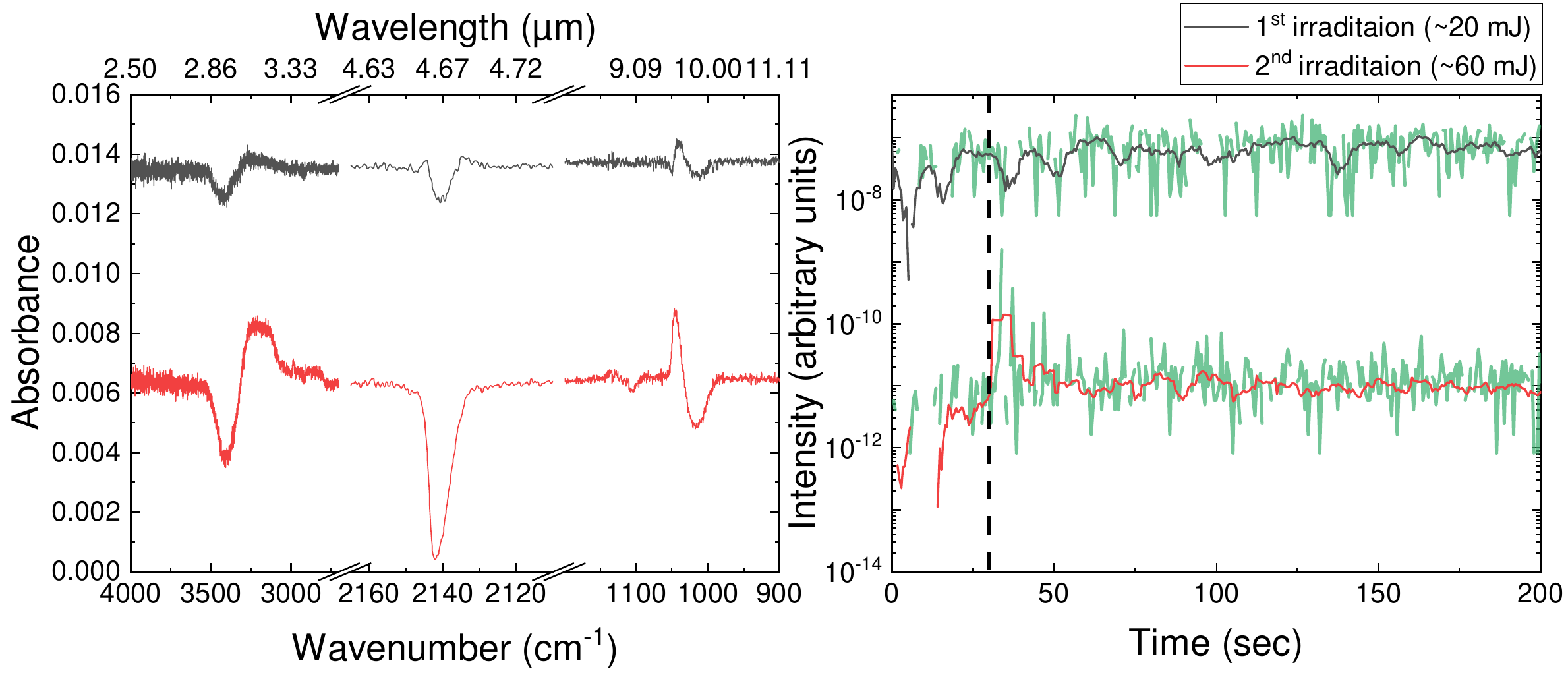}
\caption{IR and QMS data obtained for two successive 9.62 $\mu$m FEL exposures of the methanol-rich ice performed at the same substrate position (i.e., without moving the translation stage) and at 20 K. The first (black) irradiation was attenuated to a power of $\sim$20 mJ, whereas the second (red) is performed at the full attainable power of $\sim$60 mJ. Left panel: Difference of the spectra acquired before and after the irradiations. The spectra are offset for clarity. Right panel: Superposition of the QMS data for the mass signal of \ce{CO} (m/z = 28) measured concomitantly to the IR spectra during the FEL exposures (green lines) and their corresponding smoothed signal (black and red lines). The dashed line indicates the beginning of the irradiations.}
\label{fig:CO_desorb_mix}
\end{figure*}

As discussed above, the 9.62 $\mu$m difference spectrum of the methanol-rich mixture hints at the possibility of photodesorption taking place. We further investigate this hypothesis by performing successive FEL 5-minutes exposures at 9.62 $\mu$m on a same spot and varying the power settings. For the first irradiation, the $\sim$60 mJ beam is attenuated by 5 dB to yield a total power of $\sim$20 mJ. The second irradiation is performed at the full beam power of $\sim$60 mJ. The resulting difference IR spectra and corresponding QMS data are presented in Figure \ref{fig:CO_desorb_mix}.

The difference spectrum of the first irradiation is consistent with the 9.62 $\mu$m irradiation spectrum shown in the lower panel of Figure \ref{fig:mix_diff}, which was performed at similar conditions. In the case of UV experiments, photo-induced codesorption (i.e., the excitation of one species leading to the desorption of another) has been observed in the past for \ce{CO}:\ce{N2} and \ce{CO}:\ce{CO2} ice mixtures \citep{Bertin2013, Fillion2014}, but did not proceed upon the excitation of \ce{CO} in \ce{CO}:\ce{CH3OH} mixtures \citep{Bertin2016}. By contrast, both \ce{CH3OH} and \ce{CO} seem to potentially desorb from the ice to some extent upon IR excitation of methanol, given by the decrease in the intensity of their vibrational modes after the exposure to the free-electron laser. This desorption, however, is expected to be small, as no signals of \ce{CO} nor \ce{CH3OH} were detected in the QMS (see the right panel in Figure \ref{fig:CO_desorb_mix}). From the areas of the negative peaks in the IR difference spectrum, we estimate an absolute loss in column density of $\sim1.1\times10^{15}$ and $\sim0.6\times10^{15}$ molecules cm$^{-2}$ for \ce{CO} and \ce{CH3OH}, respectively. Upon increasing the beam power to its highest attainable value of $\sim$60 mJ, the second irradiation results in a difference spectrum with much more pronounced features. The \ce{CO} stretching mode decreases in intensity drastically, as shown by its prominently negative peak. Moreover, we detect an increase in the mass signal of \ce{CO} clearly above noise level upon the start of the irradiation. Comparatively, no increase in mass signal is detected for \ce{CH3OH} (not shown), and no new signals are observed in the IR spectrum. These two techniques together confirm the IR-induced photodesorption of \ce{CO} from the methanol-rich ice. The negative peaks in the IR difference spectrum after the 5-minutes irradiation yield respective \ce{CO} and \ce{CH3OH} absolute losses of $\sim8.7\times10^{15}$ and $\sim1.3\times10^{15}$ molecules cm$^{-2}$ for a fluence of $\sim9.2\times10^{22}$ photons cm$^{-2}$---corresponding to photodesorption rates of $\sim(9\pm3)\times10^{-8}$ molecules photon$^{-1}$ and $\sim(1.4\pm0.4)\times10^{-8}$ molecules photon$^{-1}$, respectively. The former is $\sim8$x higher than the estimated photodesorption rate of pure \ce{CO}, whereas the latter is $\sim2$x lower than pure \ce{CH3OH}. Since the irradiation frequency (9.62 $\mu$m) is on resonance with the $\nu_8$ vibrational mode of \ce{CH3OH}, its energy dissipation mechanisms must lead to the desorption of \ce{CO} in a more effective way than by direct vibrational excitation. This can happen, e.g., as a result of the reorganization of the methanol species towards a more stable configuration, which might spatially force \ce{CO} species out of the ice. Alternatively, energy transfer effects between interacting neighboring \ce{CO} and \ce{CH3OH} could also play a part in the sublimation of \ce{CO}---for instance through the higher density of \ce{H}-bond interactions among species in a \ce{CH3OH}-rich ice. In contrast, the photodesorption of methanol becomes less effective in the mixture than in the pure ice. These hypotheses will be investigated more deeply in future IR-FEL studies with interstellar ice analogues.

\section{Astrophysical implications}\label{sec:astro}

In this work we have focused on the infrared irradiation of interstellar ice analogues. We observe that, upon resonant vibrational irradiation, ices of pure \ce{CO}, pure \ce{CH3OH}, and mixtures of the two will generally undergo restructuring towards a more compact configuration as a result of vibrational energy dissipation. Thus, even in regions with low typical temperatures (such as dense molecular clouds), we would nonetheless expect the icy grains to contain some degree of organization---either as a result of infrared photon absorption or other phenomena leading to vibrational excitation in the ice. This could potentially influence not only the infrared band profiles of the species in the mantle, but also their diffusion, adsorption, desorption, and reaction rates---which in turn can significantly impact astrochemical models and observations. Evidently, the extent of this effect strongly depends on how localized the changes are in the ice, and how frequently it is being processed. The implications of it will therefore be increasingly intense in regions with higher radiation fluxes.

We also provide compelling evidence of the desorption of carbon monoxide induced by both vibrationally excited \ce{CO} and \ce{CH3OH} molecules, as well as the IR-induced photodesorption of pure \ce{CH3OH} ice. Those species are the main constituents of the apolar ice layer that shrouds interstellar dust grains, and therefore their photoprocesses are of great relevance to the chemistry of the interstellar medium. Given the available specific fluxes of IR photons with $\sim$4.7 and $\sim$9.6 $\mu$m inside interstellar clouds--- $\sim3\times10^{9}$ and $\sim4\times10^{8}$ photons cm$^{-2}$ s$^{-1}$, respectively \citep{Mathis1983}---and the hereby estimated upper limits to the photodesorption rates for pure \ce{aCO} and \ce{CH3OH} ices, we derive tentative IR-induced photodesorption efficiencies of $\sim3.3\times10^{1}$ molecules cm$^{-2}$ s$^{-1}$ for carbon monoxide and $\sim1.2\times10^{1}$ molecules cm$^{-2}$ s$^{-1}$ for methanol in molecular cloud conditions. Comparatively, the total UV flux (i.e., integrated for all frequencies) inside molecular clouds is estimated to be of the order of $\sim10^{4}$ photons cm$^{-2}$ s$^{-1}$ \citep{CecchiPestellini1992} which, multiplied by the derived UV-induced non-dissociative photodesorption rates of \ce{CO} and \ce{CH3OH}, yields respective efficiencies of $\sim(1.4-89)\times10^{1}$ and $\sim1.0\times10^{-1}$ molecules cm$^{-2}$ s$^{-1}$. These values are summarized in Table \ref{table:efficiencies}. Notably, the UV-induced photodesorption efficiencies for \ce{CO} are comparable to the IR-induced counterparts, and for \ce{CH3OH} they are two orders of magnitude lower. The former values are, on the one hand, presented as upper limits due to the utilization of transmission instead of reflection band strengths to derive the photodesorption rates. On the other hand, the fact that the specific IR fluxes at the species' peak absorption frequencies---instead of the integrated flux over the IR absorption width---are used in the calculations leads to an underestimation of the photodesorption efficiencies. The values listed in Table \ref{table:efficiencies} should be considered tentative, and a more thorough quantification of such efficiencies is warranted.

A caveat to these estimations is that the density of excitations (i.e., the number of simultaneously excited species in the ice) is considerably higher in laboratory experiments compared to the interstellar medium, which could play a part in the mechanisms that lead to IR-induced desorption. Still, only $\sim$0.1\% of the species in the ice are estimated to be excited per IR-FEL pulse, and therefore they should be sufficiently diluted that their interactions will have a minor effect. Moreover, the total photon fluence employed here correspond to $10^{6}-10^{7}$ yrs of exposure in interstellar conditions, thus within molecular cloud timescales. Therefore, we expect that IR-induced photodesorption can greatly affect the abundance of \ce{CO} and \ce{CH3OH} in both the solid and gas phase, and we advise to take this into consideration when modeling the chemistry of interstellar environments. In particular, IR-induced photodesorption can potentially lead to an enhanced abundance of larger species, potentially even COMs,  being preserved as they are ejected to the gas-phase from the ice, since the lower-energy IR photons are less likely to result in dissociation compared to UV rays. This could then help explain the observed abundances of \ce{CH3OH} in the gas phase. More systematic works focused on the quantification of the IR-induced photodesorption rates of interstellar ice analogues are evidently of high interest and more research is needed, both to learn about the processing taking place at a molecular level and to extrapolate these findings to interstellar environments and timescales.  

\begin{table*}[hbt!]
\centering
\caption{Comparison of the estimated fluxes, desorption rates, and desorption efficiencies of \ce{CH3OH} and \ce{CO} species induced by IR and UV photons.}
\label{table:efficiencies}      
{\begin{tabular}{lcccc}  
\toprule\midrule
&                                   &   Interstellar flux                            &   Rate                                    &   Estimated efficiency\\
&                                   &   (photons cm$^{-2}$ s$^{-1}$)    &  (molecules photon$^{-1}$)                & (molecules cm$^{-2}$ s$^{-1}$)\\
\midrule
\multirow{2}{*}{\ce{CO}}    &   IR  &   $>3\times10^{9}$ $^a$           &   $\lesssim(1.1\pm0.3)\times10^{-8}$ $^c$ &   $\sim3.3\times10^{1}$\\
                            &   UV  &   $\sim1\times10^{4}$ $^b$        &   $\sim(0.14-8.9)\times10^{-2}$ $^d$      &   $\sim(1.4-89)\times10^{1}$\\
\hline
\multirow{2}{*}{\ce{CH3OH}} &   IR  &   $>4\times10^{8}$ $^a$           &   $\lesssim(3\pm1)\times10^{-8}$ $^c$     &   $\sim1.2\times10^{1}$\\
                            &   UV  &   $\sim1\times10^{4}$ $^b$        &   $\sim1\times10^{-5}$ $^e$             &   $\sim1.0\times10^{-1}$\\

\midrule\bottomrule
\multicolumn{5}{l}{\footnotesize{$^a$ \cite{Mathis1983}.}}\\
\multicolumn{5}{l}{\footnotesize{$^b$ \cite{CecchiPestellini1992}.}}\\
\multicolumn{5}{l}{\footnotesize{$^c$ This work.}}\\
\multicolumn{5}{l}{\footnotesize{$^d$ \cite{Oberg2007, MunozCaro2010, Fayolle2011, Chen2014, Paardekooper2016}.}}\\
\multicolumn{5}{l}{\footnotesize{$^e$ \cite{Bertin2016, Cruz-Diaz2016}.}}\\
\end{tabular}}
\end{table*}

\section{Conclusions}\label{sec:conc}

In the present work, we use FELIX-2 with the LISA end station at the HFML-FELIX Laboratory to perform selective mid-infrared irradiations of \ce{CO} and \ce{CH3OH} ices, as well as mixtures of the two, under interstellar conditions. The results are monitored using both reflection-absorption infrared spectroscopy and mass spectrometry techniques. Additionally, control temperature-programmed desorption experiments on identical ices are performed to assist in the data interpretation. This study with both resonant and off-resonance infrared irradiations offers tools to elucidate the physico-chemical processes that take place in the ice as a result of vibrational energy dissipation. Our main findings are summarized below:

\begin{itemize}
    \item The vibrational excitation of the species in the ice and the subsequent dissipation of this excess energy affects the ice morphology.
    
    \item The ices explored here are restructured to a more organized configuration upon irradiation.
    
    \item In the case of pure \ce{CH3OH} ice, most of the changes in the irradiation spectra can be attributed to restructuring. However, for the pure \ce{CO} ices and mixtures with methanol, additional phenomena such as photodesorption are needed to fully explain the difference spectra.
    
    \item The changes in band shape upon irradiation are highly dependent on the FEL wavelength and the composition of the ice. Thus, the excited vibrational mode and the surrounding species must strongly affect the mechanism through which the excess vibrational energy is dissipated.
    
    \item We find compelling evidence of IR-induced photodesorption of pure \ce{CO} ice with a tentative estimated rate of $\sim(1.1\pm0.3)\times10^{-8}$ molecules photon$^{-1}$ upon excitation of its stretching mode. The photodesorption of \ce{CH3OH} is also suggested upon excitation of it $\nu_8$ mode, with a tentative estimated rate of $\sim(3\pm1)\times10^{-8}$ molecules photon$^{-1}$. Both (low) rates yield desorption efficiencies up to two orders of magnitude higher than UV-induced counterparts inside molecular clouds, because of the much higher IR fluxes in such environments. So even though the absolute rate is lower in the IR, the overall effect is expected to be around the same order of magnitude, or larger.
    
    \item Furthermore, the indirect photodesorption of \ce{CO} upon IR irradiation on resonance with \ce{CH3OH} is strongly suggested.
\end{itemize}

Given the influence of the ice morphology and composition in the rates of the processes that take place on icy grains, chemical models that involve solid-phase reactions would especially benefit from the inclusion of vibrational-energy dissipation mechanisms into the network. The tentative rates presented here will be further investigated in a follow-up study in which time is made available to derive accurate absorption cross sections for the applied reflection mode settings.
 
\begin{acknowledgements}
The authors thank the HFML-FELIX Laboratory team for their experimental assistance and scientific support, Prof. Liv Hornekær's group for sharing data for comparison, and Dr. Thanja Lamberts for the insightful discussion on the interpretation of the data presented here. The LISA end station is designed, constructed, and managed at the HFML-FELIX Laboratory by the group of S. Ioppolo and the group of B. Redlich. This work was supported by the Danish National Research Foundation through the Center of Excellence “InterCat” (Grant agreement no.: DNRF150); the Netherlands Research School for Astronomy (NOVA); the Dutch Astrochemistry Network II (DANII); the Royal Society University Research Fellowships Renewals 2019 (URF/R/191018); the Royal Society University Research Fellowship (UF130409); the Royal Society Research Fellow Enhancement Award (RGF/EA/180306); and the Royal Society Research Grant (RSG/R1/180418). Travel support was granted by the UK Engineering and Physical Sciences Research Council (UK EPSRC Grant EP/R007926/1 - FLUENCE: Felix Light for the UK: Exploiting Novel Characteristics and Expertise). K.-J.C. is grateful for support from NWO via a VENI fellowship (VI.Veni.212.296).

\end{acknowledgements}

%
%

   \bibliographystyle{aa} 
   \bibliography{mybibfile.bib} 

\begin{thebibliography}{75}
\expandafter\ifx\csname natexlab\endcsname\relax\def\natexlab#1{#1}\fi

\bibitem[{{Bertin} {et~al.}(2013){Bertin}, {Fayolle}, {Romanzin}, {Poderoso},
  {Michaut}, {Philippe}, {Jeseck}, {{\"O}berg}, {Linnartz}, \&
  {Fillion}}]{Bertin2013}
{Bertin}, M., {Fayolle}, E.~C., {Romanzin}, C., {et~al.} 2013, \apj, 779, 120

\bibitem[{{Bertin} {et~al.}(2016){Bertin}, {Romanzin}, {Doronin}, {Philippe},
  {Jeseck}, {Ligterink}, {Linnartz}, {Michaut}, \& {Fillion}}]{Bertin2016}
{Bertin}, M., {Romanzin}, C., {Doronin}, M., {et~al.} 2016, \apjl, 817, L12

\bibitem[{{Boogert} {et~al.}(2015){Boogert}, {Gerakines}, \&
  {Whittet}}]{Boogert2015}
{Boogert}, A.~C.~A., {Gerakines}, P.~A., \& {Whittet}, D. C.~B. 2015, \araa,
  53, 541

\bibitem[{{Boogert} {et~al.}(2002){Boogert}, {Hogerheijde}, \&
  {Blake}}]{Boogert2002}
{Boogert}, A.~C.~A., {Hogerheijde}, M.~R., \& {Blake}, G.~A. 2002, \apj, 568,
  761

\bibitem[{{Bottinelli} {et~al.}(2010){Bottinelli}, {Boogert}, {Bouwman},
  {Beckwith}, {van Dishoeck}, {{\"O}berg}, {Pontoppidan}, {Linnartz}, {Blake},
  {Evans}, \& {Lahuis}}]{Bottinelli2010}
{Bottinelli}, S., {Boogert}, A.~C.~A., {Bouwman}, J., {et~al.} 2010, \apj, 718,
  1100

\bibitem[{{Cecchi-Pestellini} \& {Aiello}(1992)}]{CecchiPestellini1992}
{Cecchi-Pestellini}, C. \& {Aiello}, S. 1992, \mnras, 258, 125

\bibitem[{{Charnley} {et~al.}(1992){Charnley}, {Tielens}, \&
  {Millar}}]{Charnley1992}
{Charnley}, S.~B., {Tielens}, A.~G.~G.~M., \& {Millar}, T.~J. 1992, \apjl, 399,
  L71

\bibitem[{{Chen} {et~al.}(2019){Chen}, {Lau}, {Schwarzer}, {Meyer}, {Verma}, \&
  {Wodtke}}]{Chen2019}
{Chen}, L., {Lau}, J.~A., {Schwarzer}, D., {et~al.} 2019, Science, 363, 158

\bibitem[{Chen {et~al.}(2017)Chen, Schwarzer, Verma, Stevens, Marsili, Mirin,
  Nam, \& Wodtke}]{Chen2017}
Chen, L., Schwarzer, D., Verma, V.~B., {et~al.} 2017, Accounts of Chemical
  Research, 50, 1400, pMID: 28573866

\bibitem[{{Chen} {et~al.}(2014){Chen}, {Chuang}, {Mu{\~n}oz Caro}, {Nuevo},
  {Chu}, {Yih}, {Ip}, \& {Wu}}]{Chen2014}
{Chen}, Y.~J., {Chuang}, K.~J., {Mu{\~n}oz Caro}, G.~M., {et~al.} 2014, \apj,
  781, 15

\bibitem[{{Chuang} {et~al.}(2016){Chuang}, {Fedoseev}, {Ioppolo}, {van
  Dishoeck}, \& {Linnartz}}]{Chuang2016}
{Chuang}, K.~J., {Fedoseev}, G., {Ioppolo}, S., {van Dishoeck}, E.~F., \&
  {Linnartz}, H. 2016, \mnras, 455, 1702

\bibitem[{{Chuang} {et~al.}(2018){Chuang}, {Fedoseev}, {Qasim}, {Ioppolo}, {van
  Dishoeck}, \& {Linnartz}}]{Chuang2018}
{Chuang}, K.~J., {Fedoseev}, G., {Qasim}, D., {et~al.} 2018, \apj, 853, 102

\bibitem[{{Cooke} {et~al.}(2016){Cooke}, {Fayolle}, \& {{\"O}berg}}]{Cooke2016}
{Cooke}, I.~R., {Fayolle}, E.~C., \& {{\"O}berg}, K.~I. 2016, \apj, 832, 5

\bibitem[{Corcelli \& Tully(2002)}]{Corcelli2002}
Corcelli, S.~A. \& Tully, J.~C. 2002, The Journal of Chemical Physics, 116,
  8079

\bibitem[{{Coussan} {et~al.}(2022){Coussan}, {Noble}, {Cuppen}, {Redlich}, \&
  {Ioppolo}}]{Coussan2022}
{Coussan}, S., {Noble}, J.~A., {Cuppen}, H.~M., {Redlich}, B., \& {Ioppolo}, S.
  2022, Journal of Physical Chemistry A, 126, 2262

\bibitem[{{Cruz-Diaz} {et~al.}(2016){Cruz-Diaz}, {Mart{\'\i}n-Dom{\'e}nech},
  {Mu{\~n}oz Caro}, \& {Chen}}]{Cruz-Diaz2016}
{Cruz-Diaz}, G.~A., {Mart{\'\i}n-Dom{\'e}nech}, R., {Mu{\~n}oz Caro}, G.~M., \&
  {Chen}, Y.~J. 2016, \aap, 592, A68

\bibitem[{Cuppen {et~al.}(2010)Cuppen, Ioppolo, Romanzin, \&
  Linnartz}]{Cuppen2010}
Cuppen, H.~M., Ioppolo, S., Romanzin, C., \& Linnartz, H. 2010, Phys. Chem.
  Chem. Phys., 12, 12077

\bibitem[{Cuppen {et~al.}(2022)Cuppen, Noble, Coussan, Redlich, \&
  Ioppolo}]{Cuppen2022}
Cuppen, H.~M., Noble, J.~A., Coussan, S., Redlich, B., \& Ioppolo, S. 2022, The
  Journal of Physical Chemistry A, 0, null, pMID: 36383692

\bibitem[{{Cuppen} {et~al.}(2011){Cuppen}, {Penteado}, {Isokoski}, {van der
  Marel}, \& {Linnartz}}]{Cuppen2011}
{Cuppen}, H.~M., {Penteado}, E.~M., {Isokoski}, K., {van der Marel}, N., \&
  {Linnartz}, H. 2011, \mnras, 417, 2809

\bibitem[{{Cuppen} {et~al.}(2017){Cuppen}, {Walsh}, {Lamberts}, {Semenov},
  {Garrod}, {Penteado}, \& {Ioppolo}}]{Cuppen2017}
{Cuppen}, H.~M., {Walsh}, C., {Lamberts}, T., {et~al.} 2017, \ssr, 212, 1

\bibitem[{{DeLeon} \& {Rich}(1986)}]{DeLeon1986}
{DeLeon}, R.~L. \& {Rich}, J.~W. 1986, Chemical Physics, 107, 283

\bibitem[{{DeVine} {et~al.}(2022){DeVine}, {Choudhury}, {Lau}, {Schwarzer}, \&
  {Wodtke}}]{DeVine2022}
{DeVine}, J.~A., {Choudhury}, A., {Lau}, J.~A., {Schwarzer}, D., \& {Wodtke},
  A.~M. 2022, Journal of Physical Chemistry A, 126, 2270

\bibitem[{{Draine} \& {Li}(2001)}]{Draine2001}
{Draine}, B.~T. \& {Li}, A. 2001, \apj, 551, 807

\bibitem[{{Fayolle} {et~al.}(2011){Fayolle}, {Bertin}, {Romanzin}, {Michaut},
  {{\"O}berg}, {Linnartz}, \& {Fillion}}]{Fayolle2011}
{Fayolle}, E.~C., {Bertin}, M., {Romanzin}, C., {et~al.} 2011, \apjl, 739, L36

\bibitem[{{Fayolle} {et~al.}(2013){Fayolle}, {Bertin}, {Romanzin}, {Poderoso},
  {Philippe}, {Michaut}, {Jeseck}, {Linnartz}, {{\"O}berg}, \&
  {Fillion}}]{Fayolle2013}
{Fayolle}, E.~C., {Bertin}, M., {Romanzin}, C., {et~al.} 2013, \aap, 556, A122

\bibitem[{{Fedoseev} {et~al.}(2017){Fedoseev}, {Chuang}, {Ioppolo}, {Qasim},
  {van Dishoeck}, \& {Linnartz}}]{Fedoseev2017}
{Fedoseev}, G., {Chuang}, K.~J., {Ioppolo}, S., {et~al.} 2017, \apj, 842, 52

\bibitem[{{Fillion} {et~al.}(2014){Fillion}, {Fayolle}, {Michaut}, {Doronin},
  {Philippe}, {Rakovski}, {Romanzin}, {Champion}, {{\"O}berg}, {Linnartz}, \&
  {Bertin}}]{Fillion2014}
{Fillion}, J.-H., {Fayolle}, E.~C., {Michaut}, X., {et~al.} 2014, Faraday
  Discussions, 168, 533

\bibitem[{Focsa {et~al.}(2003)Focsa, Chazallon, \& Destombes}]{Focsa2003}
Focsa, C., Chazallon, B., \& Destombes, J. 2003, Surface Science, 528, 189,
  proceedings of the Ninth International Workshop on Desorption Induced by
  Electronic Transitions

\bibitem[{{Focsa} {et~al.}(2006){Focsa}, {Mihesan}, {Ziskind}, {Chazallon},
  {Therssen}, {Desgroux}, \& {Destombes}}]{Focsa2006}
{Focsa}, C., {Mihesan}, C., {Ziskind}, M., {et~al.} 2006, Journal of Physics
  Condensed Matter, 18, S1357

\bibitem[{{Fredon} {et~al.}(2021){Fredon}, {Groenenboom}, \&
  {Cuppen}}]{Fredon2021}
{Fredon}, A., {Groenenboom}, G.~C., \& {Cuppen}, H.~M. 2021, ACS Earth and
  Space Chemistry, 5, 2032

\bibitem[{{Fuchs} {et~al.}(2009){Fuchs}, {Cuppen}, {Ioppolo}, {Romanzin},
  {Bisschop}, {Andersson}, {van Dishoeck}, \& {Linnartz}}]{Fuchs2009}
{Fuchs}, G.~W., {Cuppen}, H.~M., {Ioppolo}, S., {et~al.} 2009, \aap, 505, 629

\bibitem[{{Goumans} {et~al.}(2008){Goumans}, {Uppal}, \& {Brown}}]{Goumans2008}
{Goumans}, T.~P.~M., {Uppal}, M.~A., \& {Brown}, W.~A. 2008, \mnras, 384, 1158

\bibitem[{{Henderson} \& {Gudipati}(2014)}]{Henderson2014}
{Henderson}, B.~L. \& {Gudipati}, M.~S. 2014, Journal of Physical Chemistry A,
  118, 5454

\bibitem[{{Hiraoka} {et~al.}(1994){Hiraoka}, {Ohashi}, {Kihara}, {Yamamoto},
  {Sato}, \& {Yamashita}}]{Hiraoka1994}
{Hiraoka}, K., {Ohashi}, N., {Kihara}, Y., {et~al.} 1994, Chem. Phys. Lett.,
  229, 408

\bibitem[{{Hudgins} {et~al.}(1993){Hudgins}, {Sandford}, {Allamandola}, \&
  {Tielens}}]{Hudgins1993}
{Hudgins}, D.~M., {Sandford}, S.~A., {Allamandola}, L.~J., \& {Tielens},
  A.~G.~G.~M. 1993, \apjs, 86, 713

\bibitem[{Ioppolo {et~al.}(2010)Ioppolo, Cuppen, Romanzin, van Dishoeck, \&
  Linnartz}]{Ioppolo2010}
Ioppolo, S., Cuppen, H.~M., Romanzin, C., van Dishoeck, E.~F., \& Linnartz, H.
  2010, Phys. Chem. Chem. Phys., 12, 12065

\bibitem[{{Ioppolo} {et~al.}(2022){Ioppolo}, {Noble}, {Traspas Mui{\~n}a},
  {Cuppen}, {Coussan}, \& {Redlich}}]{Ioppolo2022}
{Ioppolo}, S., {Noble}, J.~A., {Traspas Mui{\~n}a}, A., {et~al.} 2022, Journal
  of Molecular Spectroscopy, 385, 111601

\bibitem[{{Ioppolo} {et~al.}(2011){Ioppolo}, {van Boheemen}, {Cuppen}, {van
  Dishoeck}, \& {Linnartz}}]{Ioppolo2011}
{Ioppolo}, S., {van Boheemen}, Y., {Cuppen}, H.~M., {van Dishoeck}, E.~F., \&
  {Linnartz}, H. 2011, \mnras, 413, 2281

\bibitem[{{Jakobsen} {et~al.}(1967){Jakobsen}, {Mikawa}, \&
  {Brasch}}]{Jakobsen2018}
{Jakobsen}, R.~J., {Mikawa}, Y., \& {Brasch}, J.~W. 1967, Nature, 215, 1071

\bibitem[{Jiang {et~al.}(1975)Jiang, Person, \& Brown}]{Jiang1975}
Jiang, G.~J., Person, W.~B., \& Brown, K.~G. 1975, The Journal of Chemical
  Physics, 62, 1201

\bibitem[{Kouchi(1990)}]{Kouchi1990b}
Kouchi, A. 1990, Journal of Crystal Growth, 99, 1220

\bibitem[{{Kouchi} \& {Kuroda}(1990)}]{Kouchi1990}
{Kouchi}, A. \& {Kuroda}, T. 1990, \nat, 344, 134

\bibitem[{Krasnopoler \& George(1998)}]{Krasnopoler1998}
Krasnopoler, A. \& George, S.~M. 1998, The Journal of Physical Chemistry B,
  102, 788

\bibitem[{{Li} \& {Draine}(2001)}]{Li2001}
{Li}, A. \& {Draine}, B.~T. 2001, \apj, 554, 778

\bibitem[{{Ligterink} {et~al.}(2015){Ligterink}, {Paardekooper}, {Chuang},
  {Both}, {Cruz-Diaz}, {van Helden}, \& {Linnartz}}]{Ligterink2015}
{Ligterink}, N.~F.~W., {Paardekooper}, D.~M., {Chuang}, K.~J., {et~al.} 2015,
  \aap, 584, A56

\bibitem[{Linnartz {et~al.}(2015)Linnartz, Ioppolo, \& Fedoseev}]{Linnartz2015}
Linnartz, H., Ioppolo, S., \& Fedoseev, G. 2015, International Reviews in
  Physical Chemistry, 34, 205

\bibitem[{{Luna} {et~al.}(2018){Luna}, {Molpeceres}, {Ortigoso}, {Satorre},
  {Domingo}, \& {Mat{\'e}}}]{Luna2018}
{Luna}, R., {Molpeceres}, G., {Ortigoso}, J., {et~al.} 2018, \aap, 617, A116

\bibitem[{{Mathis} {et~al.}(1983){Mathis}, {Mezger}, \& {Panagia}}]{Mathis1983}
{Mathis}, J.~S., {Mezger}, P.~G., \& {Panagia}, N. 1983, \aap, 128, 212

\bibitem[{{McClure} {et~al.}(2023){McClure}, {Rocha}, {Pontoppidan}, {Crouzet},
  {Chu}, {Dartois}, {Lamberts}, {Noble}, {Pendleton}, {Perotti}, {Qasim},
  {Rachid}, {Smith}, {Sun}, {Beck}, {Boogert}, {Brown}, {Caselli}, {Charnley},
  {Cuppen}, {Dickinson}, {Drozdovskaya}, {Egami}, {Erkal}, {Fraser}, {Garrod},
  {Harsono}, {Ioppolo}, {Jim{\'e}nez-Serra}, {Jin}, {J{\o}rgensen},
  {Kristensen}, {Lis}, {McCoustra}, {McGuire}, {Melnick}, {{\"O}berg},
  {Palumbo}, {Shimonishi}, {Sturm}, {van Dishoeck}, \&
  {Linnartz}}]{McClure2023}
{McClure}, M.~K., {Rocha}, W.~R.~M., {Pontoppidan}, K.~M., {et~al.} 2023,
  Nature Astronomy [\eprint[arXiv]{2301.09140}]

\bibitem[{{Mihesan} {et~al.}(2006){Mihesan}, {Ziskind}, {Chazallon},
  {Therssen}, {Desgroux}, {Gurlui}, \& {Focsa}}]{Mihesan2006}
{Mihesan}, C., {Ziskind}, M., {Chazallon}, B., {et~al.} 2006, Applied Surface
  Science, 253, 1090

\bibitem[{{Millar} \& {Williams}(1993)}]{Millar1993}
{Millar}, T.~J. \& {Williams}, D.~A. 1993, in Dust and Chemistry in Astronomy,
  ed. T.~J. {Millar} \& D.~A. {Williams} (Institute of Physics Publishing:
  London), 1--8

\bibitem[{{Minissale} \& {Dulieu}(2014)}]{Minissale2014}
{Minissale}, M. \& {Dulieu}, F. 2014, \jcp, 141, 014304

\bibitem[{{Minissale} {et~al.}(2016){Minissale}, {Dulieu}, {Cazaux}, \&
  {Hocuk}}]{Minissale2016}
{Minissale}, M., {Dulieu}, F., {Cazaux}, S., \& {Hocuk}, S. 2016, \aap, 585,
  A24

\bibitem[{{Mu{\~n}oz Caro} {et~al.}(2016){Mu{\~n}oz Caro}, {Chen}, {Aparicio},
  {Jim{\'e}nez-Escobar}, {Rosu-Finsen}, {Lasne}, \&
  {McCoustra}}]{Munoz-Caro2016}
{Mu{\~n}oz Caro}, G.~M., {Chen}, Y.~J., {Aparicio}, S., {et~al.} 2016, \aap,
  589, A19

\bibitem[{{Mu{\~n}oz Caro} {et~al.}(2010){Mu{\~n}oz Caro},
  {Jim{\'e}nez-Escobar}, {Mart{\'\i}n-Gago}, {Rogero}, {Atienza}, {Puertas},
  {Sobrado}, \& {Torres-Redondo}}]{MunozCaro2010}
{Mu{\~n}oz Caro}, G.~M., {Jim{\'e}nez-Escobar}, A., {Mart{\'\i}n-Gago},
  J.~{\'A}., {et~al.} 2010, \aap, 522, A108

\bibitem[{{Mu{\~n}oz Caro} \& {Mart{\'\i}n
  Dom{\'e}nech}(2018)}]{Munoz-Caro2018}
{Mu{\~n}oz Caro}, G.~M. \& {Mart{\'\i}n Dom{\'e}nech}, R. 2018, in Astrophysics
  and Space Science Library, Vol. 451, Astrophysics and Space Science Library,
  ed. G.~M. {Mu{\~n}oz Caro} \& R.~{Escribano}, 133

\bibitem[{{M{\"u}ller} {et~al.}(2021){M{\"u}ller}, {Giuliano}, {Goto}, \&
  {Caselli}}]{Muller2021}
{M{\"u}ller}, B., {Giuliano}, B.~M., {Goto}, M., \& {Caselli}, P. 2021, \aap,
  652, A126

\bibitem[{Noble {et~al.}(2020)Noble, Cuppen, Coussan, Redlich, \&
  Ioppolo}]{Noble2020}
Noble, J.~A., Cuppen, H.~M., Coussan, S., Redlich, B., \& Ioppolo, S. 2020, The
  Journal of Physical Chemistry C, 124, 20864

\bibitem[{{Noble} {et~al.}(2011){Noble}, {Dulieu}, {Congiu}, \&
  {Fraser}}]{Noble2011}
{Noble}, J.~A., {Dulieu}, F., {Congiu}, E., \& {Fraser}, H.~J. 2011, \apj, 735,
  121

\bibitem[{{Oba} {et~al.}(2018){Oba}, {Tomaru}, {Lamberts}, {Kouchi}, \&
  {Watanabe}}]{Oba2018}
{Oba}, Y., {Tomaru}, T., {Lamberts}, T., {Kouchi}, A., \& {Watanabe}, N. 2018,
  Nature Astronomy, 2, 228

\bibitem[{{Oba} {et~al.}(2010){Oba}, {Watanabe}, {Kouchi}, {Hama}, \&
  {Pirronello}}]{Oba2010}
{Oba}, Y., {Watanabe}, N., {Kouchi}, A., {Hama}, T., \& {Pirronello}, V. 2010,
  \apjl, 712, L174

\bibitem[{{{\"O}berg} {et~al.}(2007){{\"O}berg}, {Fuchs}, {Awad}, {Fraser},
  {Schlemmer}, {van Dishoeck}, \& {Linnartz}}]{Oberg2007}
{{\"O}berg}, K.~I., {Fuchs}, G.~W., {Awad}, Z., {et~al.} 2007, \apjl, 662, L23

\bibitem[{{{\"O}berg} {et~al.}(2009){{\"O}berg}, {van Dishoeck}, \&
  {Linnartz}}]{Oberg2009}
{{\"O}berg}, K.~I., {van Dishoeck}, E.~F., \& {Linnartz}, H. 2009, \aap, 496,
  281

\bibitem[{{Paardekooper} {et~al.}(2016){Paardekooper}, {Fedoseev}, {Riedo}, \&
  {Linnartz}}]{Paardekooper2016}
{Paardekooper}, D.~M., {Fedoseev}, G., {Riedo}, A., \& {Linnartz}, H. 2016,
  \aap, 596, A72

\bibitem[{{Penteado} {et~al.}(2015){Penteado}, {Boogert}, {Pontoppidan},
  {Ioppolo}, {Blake}, \& {Cuppen}}]{Penteado2015}
{Penteado}, E.~M., {Boogert}, A.~C.~A., {Pontoppidan}, K.~M., {et~al.} 2015,
  \mnras, 454, 531

\bibitem[{{Pontoppidan} {et~al.}(2003){Pontoppidan}, {Fraser}, {Dartois},
  {Thi}, {van Dishoeck}, {Boogert}, {d'Hendecourt}, {Tielens}, \&
  {Bisschop}}]{Pontoppidan2003}
{Pontoppidan}, K.~M., {Fraser}, H.~J., {Dartois}, E., {et~al.} 2003, \aap, 408,
  981

\bibitem[{{Porter} \& {Strong}(2005)}]{Porter2005}
{Porter}, T.~A. \& {Strong}, A.~W. 2005, in International Cosmic Ray
  Conference, Vol.~4, 29th International Cosmic Ray Conference (ICRC29), Volume
  4, 77

\bibitem[{{Qasim} {et~al.}(2018){Qasim}, {Chuang}, {Fedoseev}, {Ioppolo},
  {Boogert}, \& {Linnartz}}]{Qasim2018}
{Qasim}, D., {Chuang}, K.~J., {Fedoseev}, G., {et~al.} 2018, \aap, 612, A83

\bibitem[{{Roueff} {et~al.}(2014){Roueff}, {Ruaud}, {Le Petit}, {Godard}, \&
  {Le Bourlot}}]{Roueff2014}
{Roueff}, E., {Ruaud}, M., {Le Petit}, F., {Godard}, B., \& {Le Bourlot}, J.
  2014, in The Diffuse Interstellar Bands, ed. J.~{Cami} \& N.~L.~J. {Cox},
  Vol. 297, 311--320

\bibitem[{{Santos} {et~al.}(2022){Santos}, {Chuang}, {Lamberts}, {Fedoseev},
  {Ioppolo}, \& {Linnartz}}]{Santos2022}
{Santos}, J.~C., {Chuang}, K.-J., {Lamberts}, T., {et~al.} 2022, \apjl, 931,
  L33

\bibitem[{{Tielens} \& {Hagen}(1982)}]{Tielens1982}
{Tielens}, A.~G.~G.~M. \& {Hagen}, W. 1982, \aap, 114, 245

\bibitem[{{Tielens} {et~al.}(1991){Tielens}, {Tokunaga}, {Geballe}, \&
  {Baas}}]{Tielens1991}
{Tielens}, A.~G.~G.~M., {Tokunaga}, A.~T., {Geballe}, T.~R., \& {Baas}, F.
  1991, \apj, 381, 181

\bibitem[{{van Hemert} {et~al.}(2015){van Hemert}, {Takahashi}, \& {van
  Dishoeck}}]{vanHemert2015}
{van Hemert}, M.~C., {Takahashi}, J., \& {van Dishoeck}, E.~F. 2015, Journal of
  Physical Chemistry A, 119, 6354

\bibitem[{{Watanabe} \& {Kouchi}(2002)}]{Watanabe2002}
{Watanabe}, N. \& {Kouchi}, A. 2002, \apjl, 571, L173

\bibitem[{Öberg(2016)}]{Oberg2016}
Öberg, K.~I. 2016, Chemical Reviews, 116, 9631, pMID: 27099922

\end{thebibliography}

\begin{appendix}

\section{Control IR spectra}\label{appendix:control_IR}

\begin{figure*}[htb!]\centering
\includegraphics[scale=0.6]{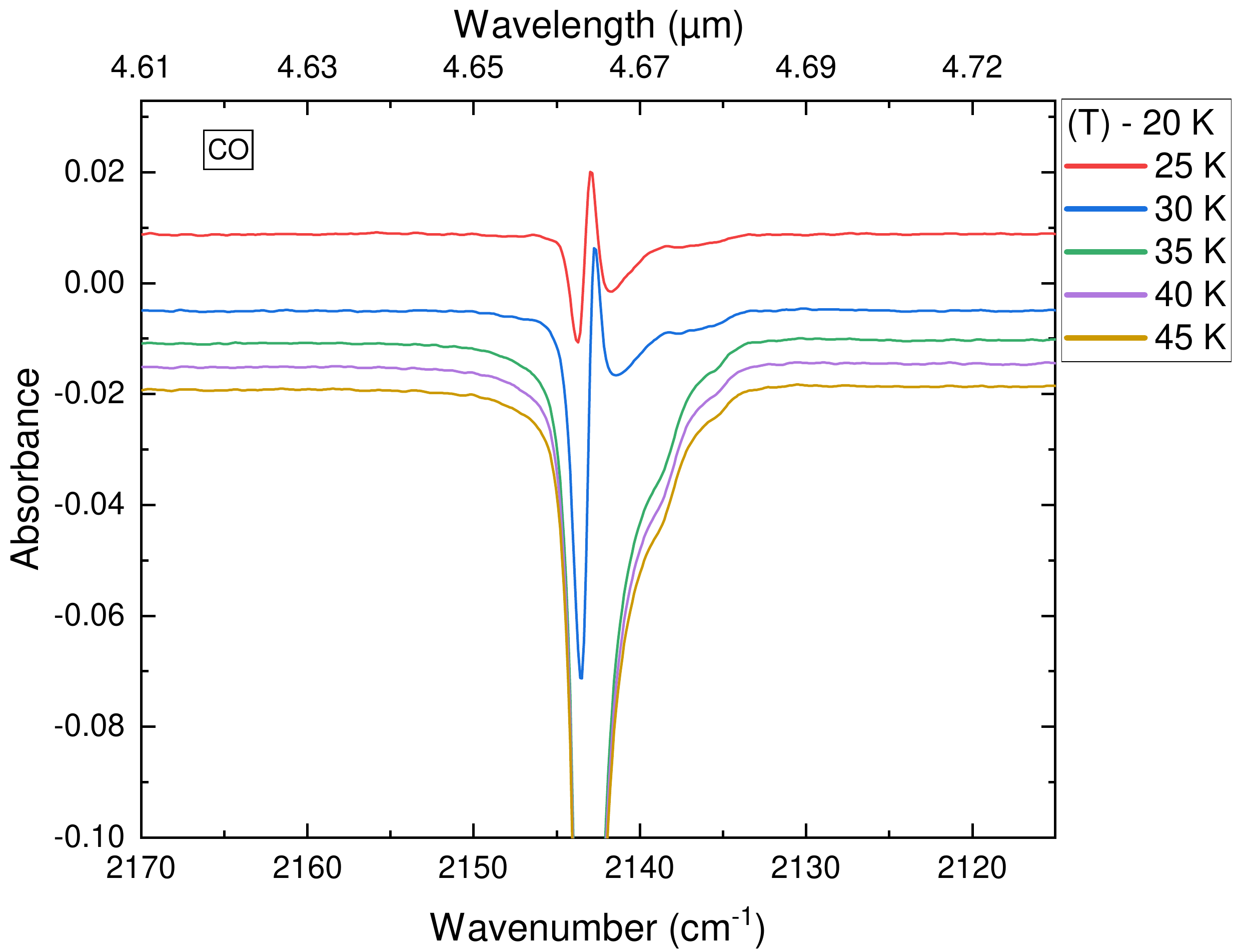}
\caption{Control IR difference spectra of pure \ce{CO} ice acquired during TPD.}
\label{fig:CO_TPD}
\end{figure*}

\begin{figure*}[htb!]\centering
\includegraphics[scale=0.6]{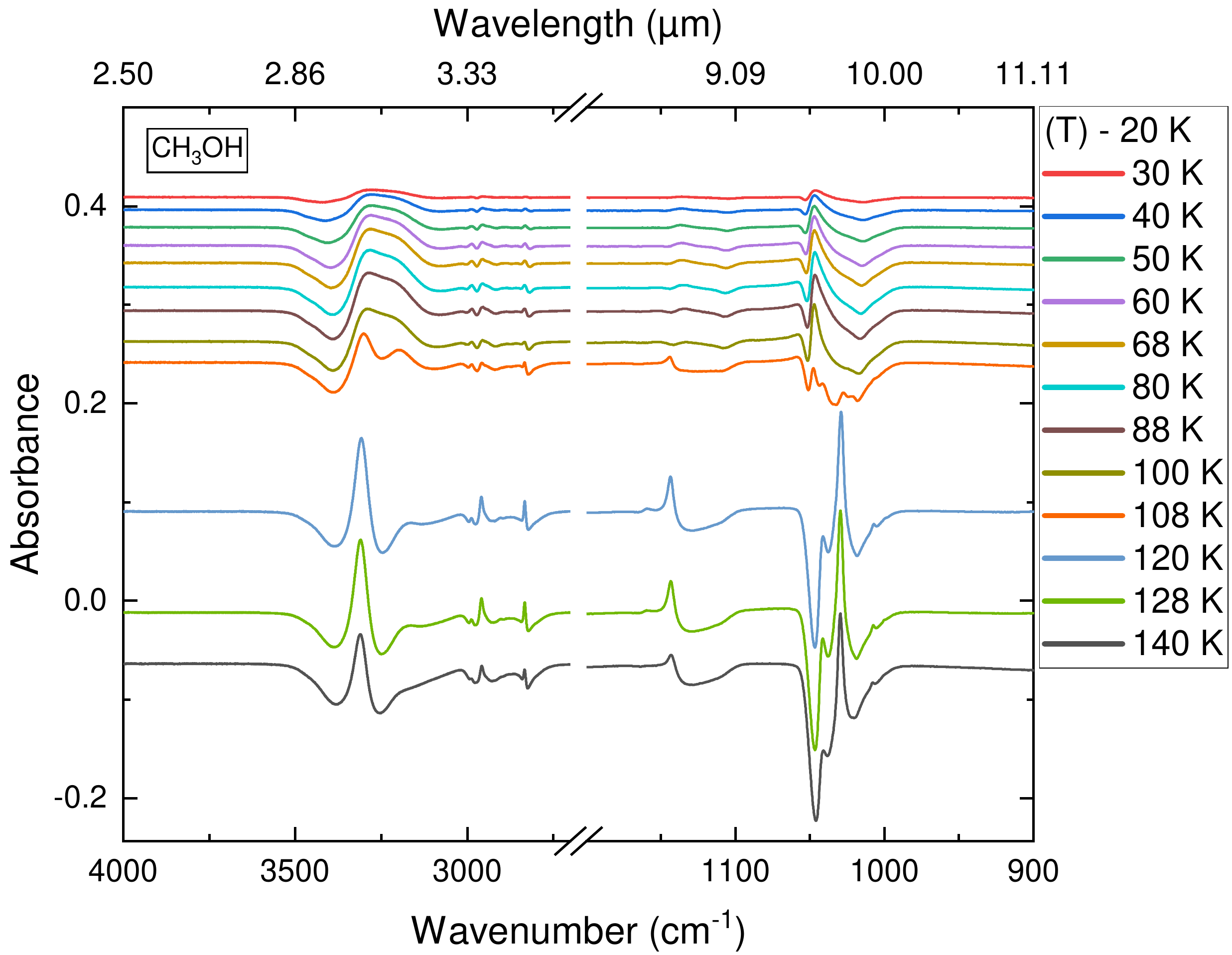}
\caption{Control IR difference spectra of pure \ce{CH3OH} ice acquired during TPD.}
\label{fig:CH3OH_TPD}
\end{figure*}

\begin{figure*}[htb!]\centering
\includegraphics[scale=0.6]{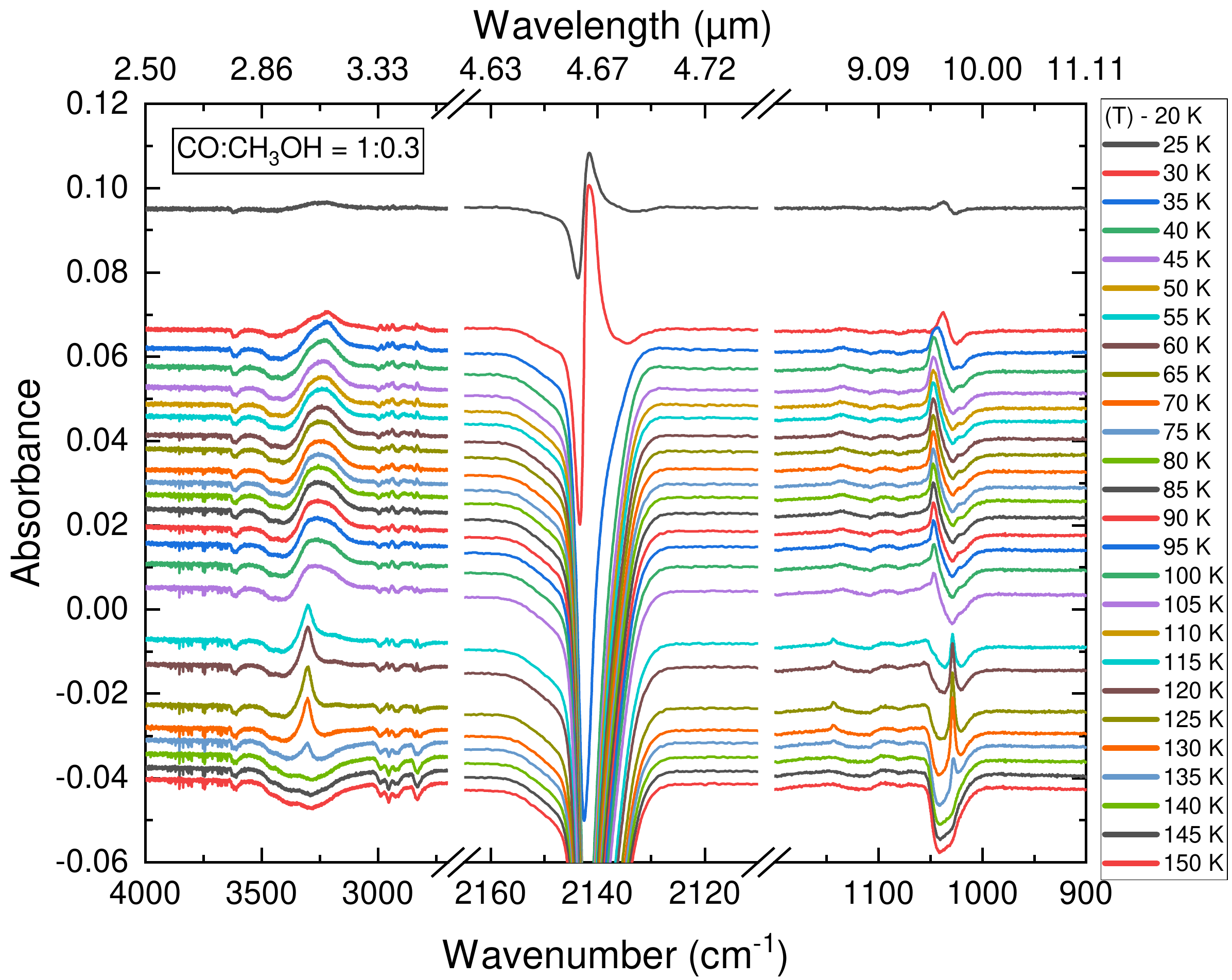}
\caption{Control IR difference spectra of the \ce{CO}:\ce{CH3OH} = 1:0.3 ice mixture acquired during TPD.}
\label{fig:COCH3OH_31_TPD}
\end{figure*}

\begin{figure*}[htb!]\centering
\includegraphics[scale=0.6]{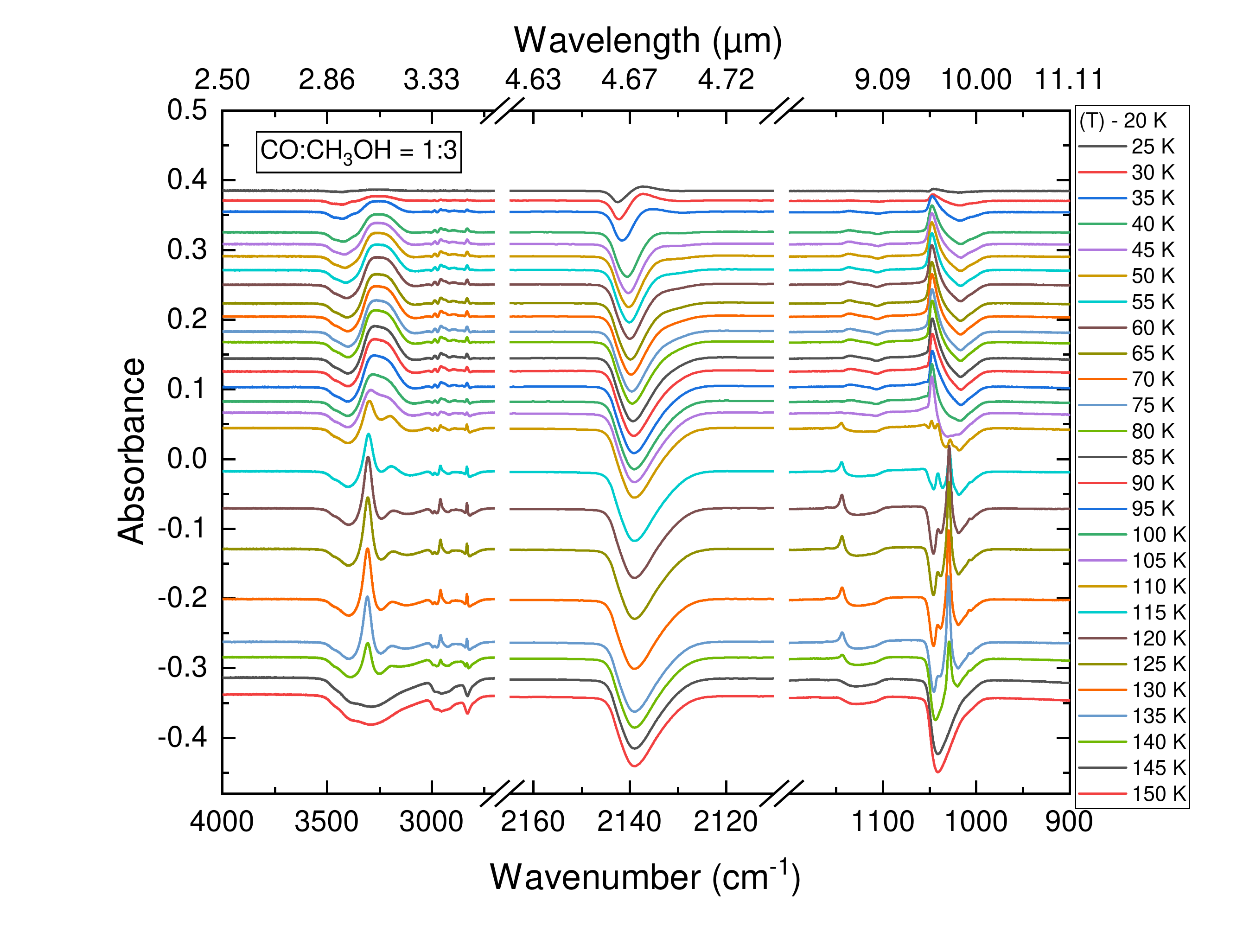}
\caption{Control IR difference spectra of the \ce{CO}:\ce{CH3OH} = 1:3 ice mixture acquired during TPD.}
\label{fig:COCH3OH_13_TPD}
\end{figure*}

The difference IR spectra of the \ce{CO}, \ce{CH3OH}, \ce{CO}:\ce{CH3OH} = 1:0.3, and \ce{CO}:\ce{CH3OH} = 1:3 ices acquired during the control TPD experiments are shown in Figures \ref{fig:CO_TPD}, \ref{fig:CH3OH_TPD}, \ref{fig:COCH3OH_31_TPD}, and \ref{fig:COCH3OH_13_TPD}, respectively. Differences are taken between a temperature T (as labeled in the legend) and the deposition temperature of 20 K.

\section{\ce{aCO} ice stabilization}\label{appendix:co_stab}

\begin{figure*}[htb!]\centering
\includegraphics[scale=0.5]{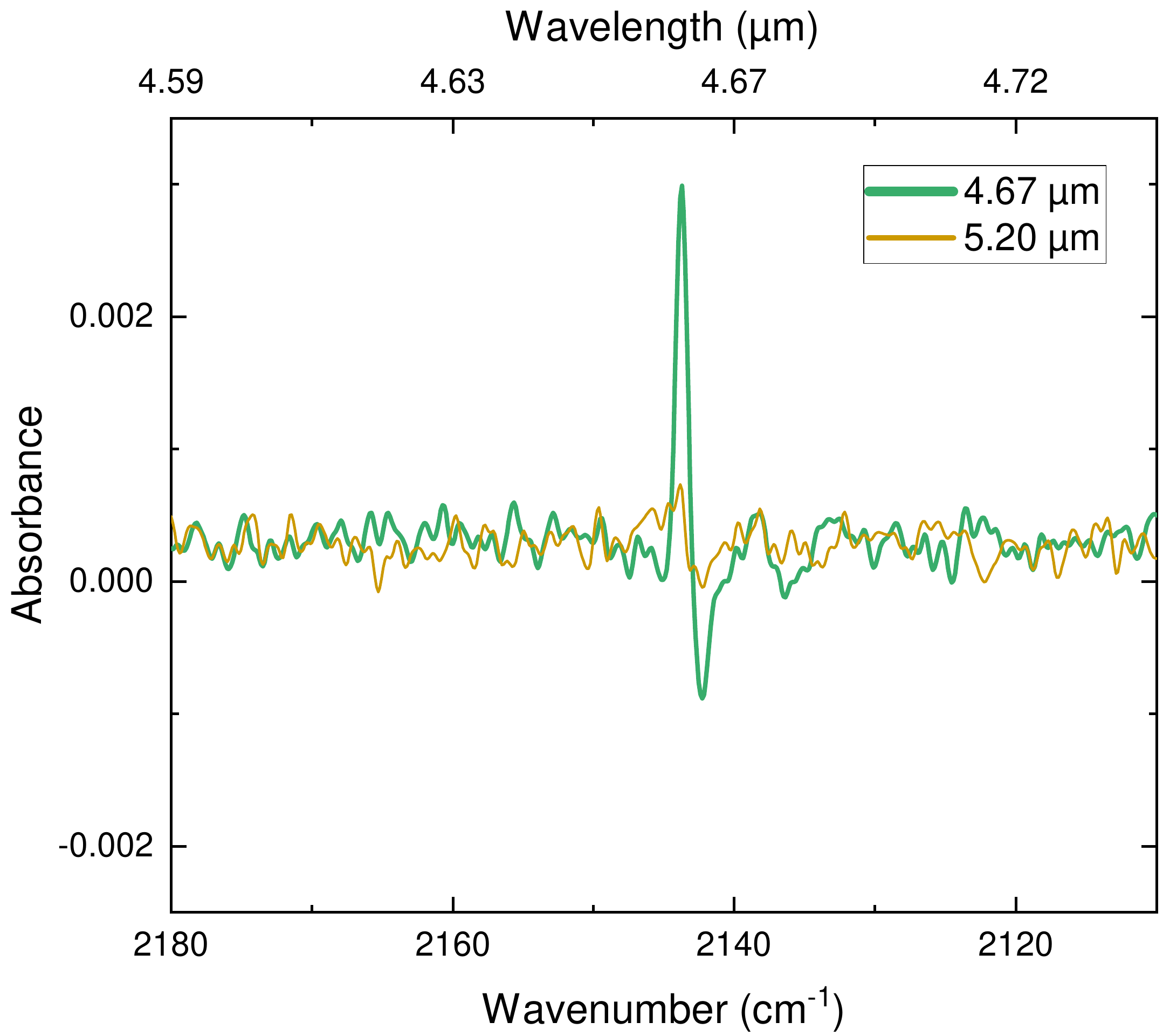}
\caption{Difference spectra obtained before and after 10-minutes IR-FEL irradiation on aCO ice at 20 K, corrected for ice stabilization.}
\label{fig:CO_stab}
\end{figure*}

The difference IR spectra before and after irradiation of \ce{aCO} ice corrected for stabilization are shown in Figure \ref{fig:CO_stab}. The correction is made by subtracting the signal due to ice stabilization---as obtained from two spectra taken 10 minutes apart and without IR-FEL exposure---from the irradiation difference spectra.

\section{Repeated irradiations on ice mixtures}\label{appendix:repeated_mix}

\begin{figure*}[htb!]\centering
\includegraphics[scale=0.6]{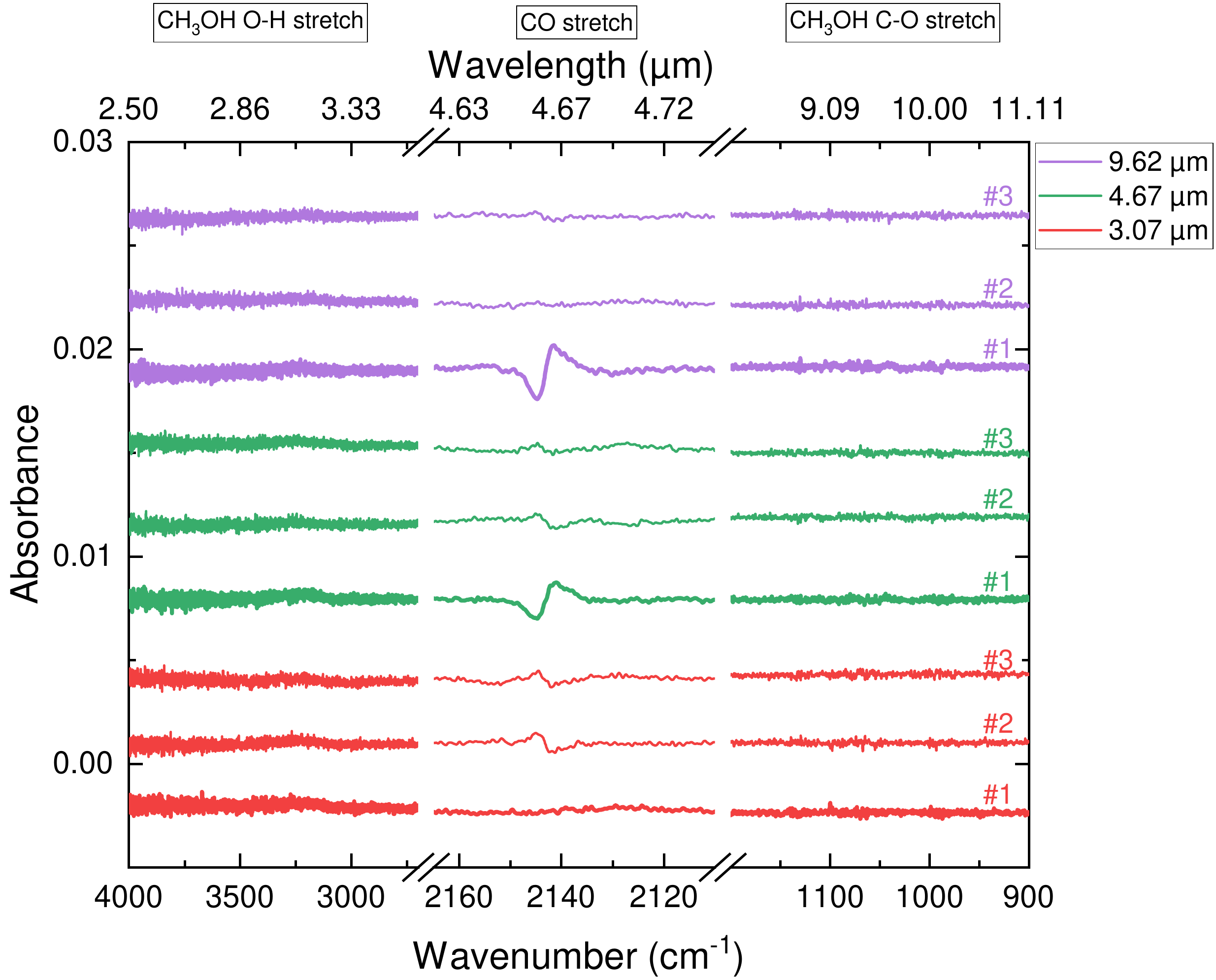}
\caption{Difference irradiation spectra of repeated measurements on the \ce{CO}-rich ice mixture. Irradiations at the same frequency were performed at the same substrate position, and the ice is already saturated on each second and third measurement.}
\label{fig:CO_rich_app}
\end{figure*}

\begin{figure*}[htb!]\centering
\includegraphics[scale=0.6]{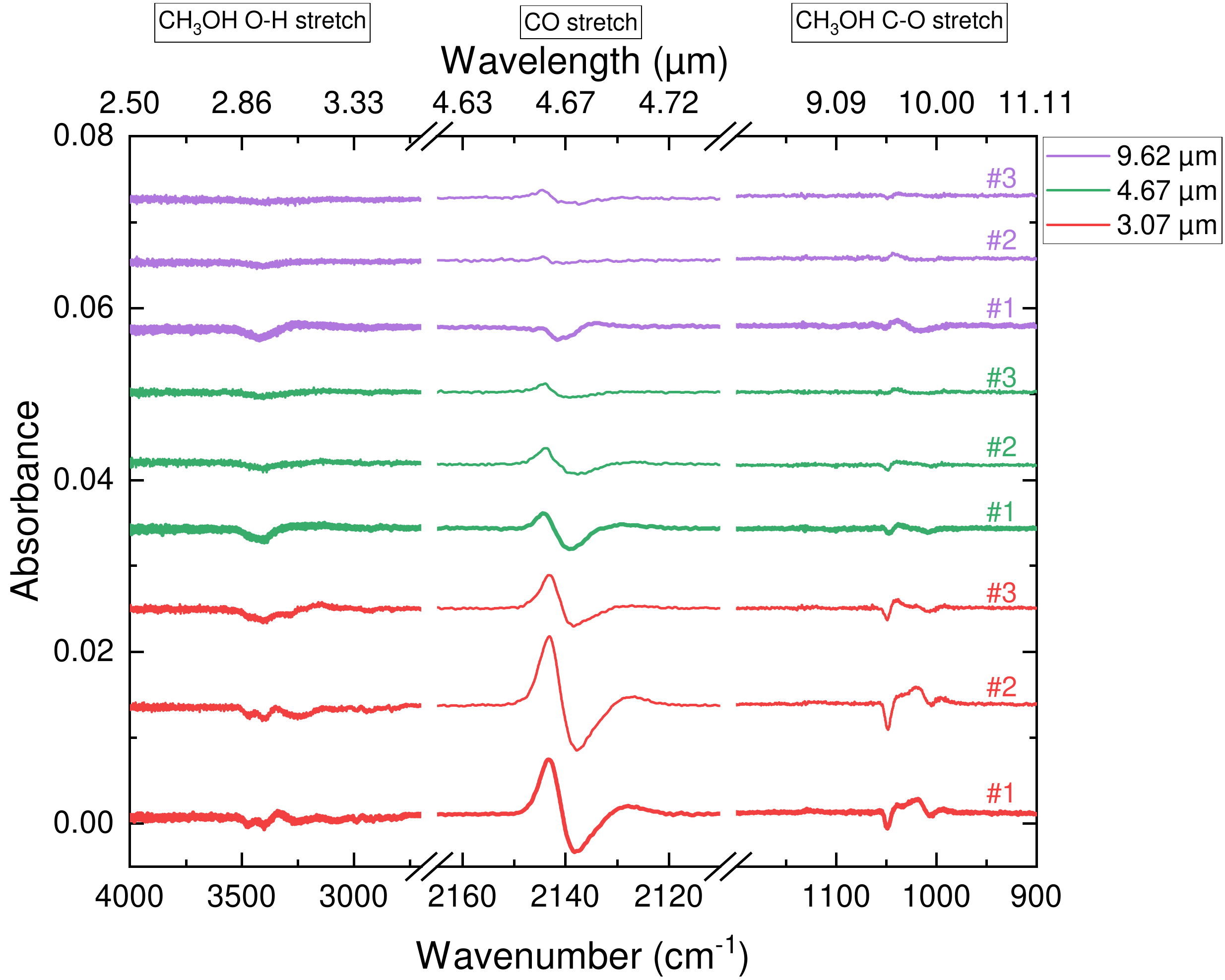}
\caption{Difference irradiation spectra of repeated measurements on the \ce{CH3OH}-rich ice mixture. Irradiations at the same frequency were performed at the same substrate position.}
\label{fig:CH3OH_rich_app}
\end{figure*}

The irradiations of the ice mixtures were performed a total of three times in a same substrate position each, i.e., the translator stage was only used when the irradiation wavelength was changed. The difference spectra of the \ce{CO}-rich and \ce{CH3OH}-rich ice mixtures are shown in Figures \ref{fig:CO_rich_app} and \ref{fig:CH3OH_rich_app}, respectively. In the \ce{CO}-rich case, the ice becomes saturated after the first irradiation at each frequency, and no significant effect---besides ice stabilization---can be seen in the repeated measurements. The irradiation at 9.62 $\mu$m was performed significantly later in the experiment shift, in which case the ice had had appreciably more time to stabilize---hence the smaller signal in its second and third measurements. In the \ce{CH3OH}-rich ice case, saturation takes higher fluences to occur, and the effects of the IR-FEL are still observable in the repeated irradiations.

\end{appendix}

\end{document}